\documentclass{article}
\usepackage{arxiv}
\usepackage{multirow}
\usepackage[utf8]{inputenc} % allow utf-8 input
\usepackage[T1]{fontenc}    % use 8-bit T1 fonts
\usepackage{hyperref}       % hyperlinks
\usepackage{url}            % simple URL typesetting
\usepackage{booktabs}       % professional-quality tables
\usepackage{amsfonts}       % blackboard math symbols
\usepackage{nicefrac}       % compact symbols for 1/2, etc.
\usepackage{microtype}      % microtypography
\usepackage{lipsum}		% Can be removed after putting your text content
\usepackage{graphicx}
\usepackage[percent]{overpic}
\usepackage{subcaption}
\usepackage{natbib}
\usepackage{doi}
\usepackage{amsmath}
\usepackage{longtable}
\usepackage[table, dvipsnames]{xcolor}
\usepackage{amssymb}  % for \checkmark
\renewcommand{\arraystretch}{1.3}

\hypersetup{
    colorlinks=true,  % Enable colored links (not boxed)
    linkcolor=RoyalBlue,   % Color of internal links
    urlcolor=RoyalBlue,    % Color of URL links
    citecolor=OliveGreen  % Color of citations
}

\usepackage{tcolorbox}

\newtcolorbox{metricbox}[1]{
    colback=gray!10,
    colframe=gray!50,
    arc=3mm,
    boxrule=0.5pt,
    title=#1,
    fonttitle=\bfseries,
    coltitle=black,
    colbacktitle=gray!20,
    before skip=10pt,
    after skip=10pt
}

\title{Politics and polarization on Bluesky}

\author{ \href{https://orcid.org/0000-0002-2381-6876}{\includegraphics[scale=0.06]{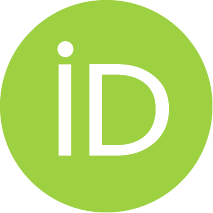}\hspace{1mm} Ali Salloum} \\
	Department of Computer Science\\
	Aalto University\\
	\texttt{ali.salloum@aalto.fi} \\
    \And
	\href{https://orcid.org/0000-0001-8257-7214}{\includegraphics[scale=0.06]{orcid.pdf}\hspace{1mm}Dorian Quelle} \\
    Department of Mathematical Modeling and Machine Learning \\
    Digital Society Initiative \\
    University of Zurich \\
    \texttt{dorian.quelle@uzh.ch}
	\And
	\href{https://orcid.org/0000-0002-0635-9933}{\includegraphics[scale=0.06]{orcid.pdf}\hspace{1mm}Letizia Iannucci} \\
	Department of Computer Science\\
	Aalto University\\
	\texttt{letizia.iannucci@aalto.fi} \\
    \And
    \href{https://orcid.org/0000-0003-3937-3704}{\includegraphics[scale=0.06]{orcid.pdf}\hspace{1mm}Alexandre Bovet} \\
    Department of Mathematical Modeling and Machine Learning \\
    Digital Society Initiative \\
    University of Zurich \\
    \texttt{alexandre.bovet@uzh.ch}
    \And
	\href{https://orcid.org/0000-0003-2049-1954}{\includegraphics[scale=0.06]{orcid.pdf}\hspace{1mm}Mikko Kivelä} \\
	Department of Computer Science\\
	Aalto University\\
	\texttt{mikko.kivela@aalto.fi} \\
}

\renewcommand{\headeright}{}
\renewcommand{\undertitle}{}
\renewcommand{\shorttitle}{\textit{arXiv} Template}

\hypersetup{
pdftitle={A template for the arxiv style},
pdfsubject={q-bio.NC, q-bio.QM},
pdfauthor={David S.~Hippocampus, Elias D.~Striatum},
pdfkeywords={First keyword, Second keyword, More},
}

\renewcommand{\arraystretch}{1}

\begin{document}
\maketitle

\begin{abstract}

Online political discourse is increasingly shaped not by a few dominant platforms but by a fragmented ecosystem of social media spaces---each with its own user base, target audience, and algorithmic mediation of discussion. Such fragmentation may fundamentally change how polarization manifests online. In this study, we investigate the characteristics of political discourse and polarization on the emerging social media site Bluesky. We collect all activity on the platform between December 2024 and May 2025 to map out the platform's political topic landscape and detect distinct polarization patterns. Our comprehensive data collection allows us to employ a data-driven methodology for identifying political themes, classifying user stances, and measuring both structural and content-based polarization across key topics raised in English-language discussions. Our analysis reveals that approximately 13\% of Bluesky posts engage with political content, with prominent topics including international conflicts, U.S. politics, and socio-technological debates. We find high levels of structural polarization across several salient political topics. However, the most polarized topics are also  highly imbalanced in the numbers of users on opposing sides, with the smaller group consisting of only 1--2\% of the users. While discussions in Bluesky echo familiar political narratives and polarization trends, the platform exhibits a more politically homogeneous user base than was typical prior to the current wave of platform fragmentation.

\end{abstract}

% keywords can be removed

\section{Introduction}

The rise of new digital platforms substantially changed the ways in which people communicate about politics. As these platforms become central venues for public discussion and information sharing, understanding the dynamics of political conversations within them has become an important area of study. While much attention has been paid to established platforms such as Twitter \cite{conover2011political} and Facebook \cite{del2016echo}, comparatively few studies have investigated emerging platforms like Bluesky. Specifically, little is known about what topics are discussed and whether patterns of polarization mirror those on more established social media sites \cite{kleppmann2024bluesky}.

To ensure validity when researching novel social media sites, it is crucial to avoid making strong assumptions about which topics will be relevant or how polarization will manifest. Many previous studies have approached this issue by focusing on pre-selected topics or well-known partisan markers, which can limit the discovery of new or unexpected trends. By minimizing these assumptions and allowing the data to reveal the topics discussed and the nature of any divisions, researchers can gain a more accurate picture of the communication taking place on the platform.

This paper adopts an open, data-driven approach to explore the main topics of discussion on Bluesky, and to examine the presence and characteristics of polarization within these conversations. In doing so, it seeks to provide an impartial view of political communication on this platform, and to contribute to a broader understanding of how political discourse evolves online. To guide our analysis, we pose two research questions: (1) What political topics are discussed on Bluesky? (2) Do these discussions show signs of political polarization?

Studying these questions is important for several reasons. First, decentralized platforms like Bluesky have not yet been widely examined in the scholarly literature, despite their growing influence. Second, Bluesky is providing unprecedented and unimpeded data access for researchers, enabling the observation of richer interactions among the users. Third, these online spaces serve as a window into the issues and attitudes that matter most to users, which do not always align with the agendas set by politicians or established media \cite{gilens2009preference, king2017news}. Finally, as digital spaces increasingly shape political attitudes and behaviors, it is important to better understand how discussions and divisions emerge in these new environments \cite{cobbe2021algorithmic}. 

The remainder of this paper is structured as follows. In Section \ref{sec:methods}, we detail our methodology, including data collection, topic modeling, stance classification, and network construction. We also describe the analytical measures and group detection techniques used to assess polarization. Section \ref{sec:findings} presents our findings, beginning with an overview of activity patterns and topical diversity on Bluesky, followed by a focused analysis of political content and the structure of polarization within and across key topics. Finally, in Section \ref{sec:conclusion}, we discuss the broader implications of our results, compare Bluesky’s dynamics to those observed on other platforms, and reflect on the limitations and future directions for this line of research.

\section{Methods}
\label{sec:methods}
\begin{figure}[t]
    \centering
    \includegraphics[width=\textwidth]{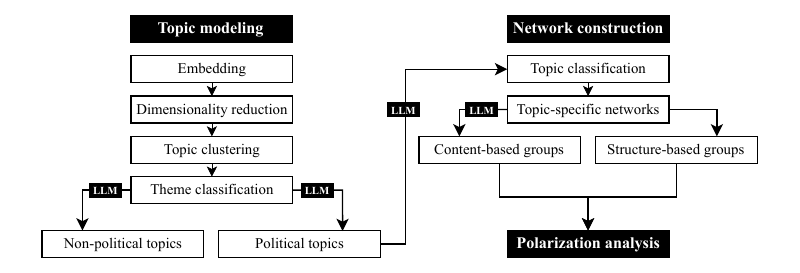}
    \caption{\textbf{Analytical pipeline for topic detection and network construction.} Overview of our analytical pipeline, which contains three main stages: (1) We map the landscape of discussion on the platform by using embeddings, dimensionality reduction, and clustering to identify coherent groups of posts; an LLM then classifies clusters as political or non-political. (2) All posts and interactions labeled as political are further categorized by the LLM into specific political topics, and for each topic, a directed user network is constructed, where nodes represent users who have posted or reposted relevant content. These topic-based networks are then clustered by structural or content-based features. (3) Finally, these networks serve as the basis for downstream analyses, with a focus on measuring and understanding polarization on the platform.}
    \label{fig:pipeline_schema}
\end{figure}

This section is organized into five subsections. The first describes the data collection process. The second details our approach to topic modeling, including how topics of conversation were identified and how political topics were distinguished from non-political ones. The third subsection explains how topics were transformed into networks. The fourth introduces two methods used to identify functional groups, which helps us further understand the network structure. Finally, the fifth subsection lists the measures computed for the networks to describe their intrinsic properties, and to evaluate the level of polarization in the discussions. A visual overview of our pipeline is provided in Fig.~\ref{fig:pipeline_schema}.

\subsection{Data collection}
For this study, we collected all interaction data from Bluesky spanning the period from December 2024 to May 2025 using the Bluesky Firehose API \cite{bluesky_firehose}. The dataset includes events related to various types of user activities and interactions on the platform. Each event in the dataset is characterized by its category: \textit{create}, \textit{update}, or \textit{delete}. For our analysis, we specifically focused on \textit{create} actions, and on the native Bluesky event types \textit{app.bsky.feed.post} (publishing a new post), \textit{app.bsky.feed.repost} (resharing an existing post), \textit{app.bsky.feed.like} (liking an existing post), \textit{app.bsky.graph.block} (blocking another user), and
\textit{app.bsky.graph.follow} (starting to follow another user). Our collection covers all languages and includes every user who interacted in any form from 17 December 2024 (when our streaming data collection started) onwards. We maintained a data collection uptime of over 98\% (see Appendix~\ref{appendix-data-collection}). 

\subsection{Topic modeling}
To extract the distribution of topics that are discussed on Bluesky, we first queried the raw data for all posts that had received at least one repost. To reduce noise, we retained only those posts containing at least five characters. Since our preliminary analysis showed that over 80\% of the posts were in English, we focused our initial study on English-language content, using the language tags inferred by Bluesky. %While these tags are not perfect, they were sufficiently reliable for our purposes.

\paragraph{Identifying semantic clusters}
\label{secsec:finding-topics}
From this filtered dataset, we randomly sampled 3\% of the posts for the overall topic exploration. Each post was then embedded using the \textit{E5-mistral-7b-instruct model} \cite{wang2023improving}, which produces 4096-dimensional embeddings and is among the top performers on the MTEB leaderboard \cite{muennighoff2022mteb, mteb_leaderboard}. We selected this model because its embeddings produced clearer, more nuanced clusters compared to three smaller embedding models we tested \textit{(all-MiniLM-L6/L12-v2 \& 
all-mpnet-base-v2)}.

Given the high dimensionality of these embeddings, we applied UMAP \cite{McInnes2018} to reduce the vectors to a 128-dimensional space. We then used HDBSCAN \cite{10.5555/3001460.3001507} to cluster the posts within this reduced space. For each cluster, we selected a sample of 50 posts and used a large language model, 
the open source \textit{gemma-3-12b-it} (Gemma 12B) \cite{team2025gemma}, to generate summaries of the matters discussed in these posts, which we then reviewed manually. We will use these data-driven clusters as a starting point for refining and selecting the final topics.

%We refer to these topic clusters as \textit{subtopics} in this study.

\subsection{Topic classification}
\paragraph{Political semantic clusters}
As this study focuses on political discussions, each semantic cluster must be classified as either political or non-political. Perspectives on a given subtopic can differ significantly. For example, the hypothetical cluster of \emph{Carbon Emissions Regulation} spans several themes. As an Infrastructure \& Environment issue, it relates to environmental protection and sustainable development. Within Economy, Trade \& Labor, it concerns regulations, green jobs, and industry costs. From the perspective of Science, Technology \& Energy, debates revolve around scientific evidence, innovation, and energy transitions. 

To assess the concentration of political content within each identified cluster, and to determine which political themes are most prevalent (if any), we use the large language model \textit{mistral-Small-3.1-24B-Instruct-2503} (Mistral 24B) for classification. The themes applied in this study are adapted from the \textit{Comparative Agendas Project} (CAP) Codebook \cite{jones2023policy}, which offers a comprehensive and widely recognized taxonomy for categorizing political topics across diverse policy areas (see the selected themes and their descriptions in Appendix Table \ref{themetable}). We prompted the LLM to assign each post to the most appropriate category, including a ``Non-Political'' label if a post did not fit any political theme. For each semantic cluster, this produces a distribution of political themes across the posts associated with it. Finally, if a semantic clusters consists of 75\% or more posts labeled as ``Non-Political'', we classified it as apolitical; otherwise, it was considered a political semantic clusters. The full list of inferred semantic clusters, along with their classification, is provided in Appendix Table \ref{appendix-subtopics}.

We applied the aforementioned methodology to the totality of posts on Bluesky, including those not belonging to any semantic cluster, to obtain a comprehensive understanding of prevalent political themes on Bluesky. The overall prevalence numbers are reported in Table \ref{tab:post_distribution}.

\paragraph{Political topics}
\label{parent-topics}
To assure that the semantic clusters are meaninful and not duplicated, we manually define $k$ topics that best represent the semantic clusters. This ensures that topics are interpretable. In the following, we refer to these as \textit{political topics}. We reclassify all posts labeled as political into the $k$ parent topics, ensuring that the content not the semantics determines the posts assignment. This step is crucial because different groups may use distinct vocabularies or rhetorical styles when discussing the same issue, meaning that relying solely on semantic similarity could overlook important connections. We used \emph{Mistral 24B} to assign each political post to the most appropriate political topic. We added an ``Other'' category to prevent posts from being arbitrarily assigned to political topics when they don't clearly fit the predefined categories. Hereafter, these \textit{political topics} will be used to investigate patterns of polarization further.

\subsection{Network construction}
\label{network_construction}

Now that posts have been classified by political topic, we can construct interaction graphs for each topic to capture how individuals engage with one another around specific issues. Our approach starts with a bipartite network, where users and posts are represented as two distinct node types, connected by user interactions with posts. To examine relationships between users within each political topic, we project this bipartite network onto the user dimension, transforming indirect, post-mediated interactions into direct user-to-user connections. In this projected network, each time a user reposts another user’s content, a parallel edge is created between them.

To formalize this setup, we introduce notation for the topic- and time-specific networks constructed in our analysis. Let $T$ denote the set of discrete time points (e.g., months), and let $X$ represent the set of identified political topics (e.g., Climate politics). For each topic $x \in X$ and time point $t \in T$, we define a family of directed graphs $G^{\tau}_{x, t} = (V^{\tau}_{x, t}, E^{\tau}_{x, t})$, where $\tau \in \{ \mathrm{likes},\ \mathrm{reposts},\ \mathrm{follows},\ \mathrm{blocks} \}$ indexes the interaction type. The node set $V^{\tau}_{x, t}$ consists of users participating in topic $x$ at time $t$ through type $\tau$, and each edge in $E^{\tau}_{x, t}$ represents an observed directed interaction between two users.

In this study, for each political topic, we define $\tau$ as \textit{reposts} and set $T$ as the single time period from December 2024 to May 2025, focusing our analysis on one repost network per political topic.

Note, however, while our main analysis focuses on repost interactions, this framework can be readily extended to include other types of user interactions. Unlike reposts and likes, follows and blocks relate directly to user-to-user relationships rather than user-content interactions. For a given topic $x$, one can take the union of user sets from the likes and reposts networks, $V^{\text{likes}}_{x, t} \cup V^{\text{reposts}}_{x, t}$, and construct follower and blocking networks on this shared set of users. In general, this allows the definition of a tuple of interaction networks for each $(x,t)$ pair: $\mathbf{G}_{x,t} = \left( G_{x,t}^{\text{likes}},\ G_{x,t}^{\text{reposts}},\ G_{x,t}^{\text{follows}},\ G_{x,t}^{\text{blocks}}\right)$. The collection of all such tuples across topics and time points, $\mathbf{G} = \left\{ \mathbf{G}_{x,t} : x \in X,\ t \in T \right\}$, provides a unified framework for analyzing multiple interaction types.

\subsection{Group detection}
\label{group-detection}
The study of polarization in social networks often begins with identifying distinct groups, since polarization is inherently a group-level phenomenon \cite{chen2021polarization}. These groups may be revealed through structural analysis of interaction patterns \cite{salloum2022separating, garimella2018quantifying}, or by examining the content that users generate and share.

In highly polarized contexts, prior studies have repeatedly found that polarization often appears as tightly clustered communities \cite{conover2011political, garimella2018quantifying, salloum2022separating, salloum2024anatomy}, typically forming two main groups that represent opposing perspectives. However, the presence of modular structure in a network does not necessarily imply polarization. Subcommunities may also arise among individuals who share similar views, and platform-specific algorithms can shape group boundaries by curating users’ feeds and limiting exposure to posts, even among those with aligned positions.

To address that, researchers typically manually inspect samples to verify that the detected structural groups correspond to actual opposing political camps for the polarization analysis. Alternatively, group boundaries can be inferred from content analysis. Here, the use of large language models presents a promising approach, enabling scalable and increasingly accurate annotation of group alignments based on textual data.

In this study, we employ two complementary approaches to determine user stances on each identified political topic: \textit{content-based group detection} and \textit{structure-based group detection}. By combining these methods, we aim to take advantage of the strengths of both perspectives, capturing not only what users say, but also how they connect.

\paragraph{Content-based group detection}
\label{content-based-detection}

For each political topic, we identify users who have participated in the relevant discussion. To assess their stance on the topic, we prompt the LLM (Mistral 24B) to evaluate each user's position. For every user, we sample up to ten posts that they have either authored or reposted, and provide these posts as context in the LLM prompt. The model then assigns a stance label--for, neutral, or against--to each user, based on the sampled content.

\paragraph{Structure-based group detection}
\label{structural-based-detection}
In addition to content analysis, we apply a structural approach \cite{salloum2022separating} using the stochastic block model \cite{peixoto2017nonparametric}, specifically the planted partition model \cite{zhang2020statistical}, to detect groups within the network. We set the maximum number of groups to five and perform 15 independent runs of the model, each with 50 iterations, selecting the partition with the lowest description length -- according to the \textit{Occam's razor principle}. For each resulting group, we examine the distribution of content-based stances among its members.

\subsection{Political polarization}
\label{political-polarization-section}

In addition to some foundational metrics, we report, for each network, a range of polarization-related scores, including measures of homophily (the extent to which similar users are connected), division (the degree of separation or clustering among groups within the network), and group size imbalance (disparities in the sizes of opposing or distinct groups). The following scores are reported:

\begin{metricbox}{Metrics for Individual Networks}
\textbf{Adaptive EI Index} \cite{chen2021polarization} quantifies structural polarization in networks by comparing the density of connections within groups to the density of connections between groups. A higher AEI score indicates that connections are more concentrated within groups than between them, reflecting a potentially more polarized network structure.  \\

\textbf{Assortativity Coefficient} \cite{newman2003mixing} measures the tendency of nodes in a network to connect with others that are similar in some way, such as belonging to the same group or category. A higher assortativity value indicates that nodes are more likely to be connected to others of the same type, reflecting strong homophily within the network.  \\

\textbf{Coleman Homophily Index} \cite{coleman1958relational} quantifies the degree to which members of a particular group preferentially form connections within their own group, relative to what would be expected by chance. A value of zero indicates random mixing, positive values reflect a tendency toward within-group connections (homophily), and negative values indicate a preference for connections with other groups (heterophily). The index provides a normalized measure, making it possible to compare homophily levels across groups of different sizes.  \\

\textbf{Simpson’s Diversity Index} \cite{simpson1949measurement} measures the diversity of group membership within a network or a subset of nodes, quantifying the probability that two randomly selected individuals belong to different groups. A higher value indicates greater diversity (i.e., a more even distribution among groups), while a lower value suggests dominance by a single group. 
\end{metricbox} 

While it is useful to study each network separately, additional insights can be gained by considering how topics and user groups overlap. Previous research has shown that certain topics often attract the same users, and that individuals’ stances may be correlated across multiple topics, even when those topics are not directly related \cite{chen2021polarization, salloum2024anatomy}. To explore these broader patterns in political communication, we also report two scores that capture these types of dynamics.

\begin{metricbox}{Metrics for Multiple Networks}
\textbf{Jaccard Index} is used to evaluate the overlap between topics by measuring the proportion of shared users (nodes) between pairs of topic networks relative to their combined unique users. In addition, we construct a hypergraph where topics are represented as nodes and hyperedges are formed for bundle of topics whose degree of overlap exceeds a specified threshold. \\

\textbf{Issue alignment} \cite{chen2021polarization} quantifies the extent to which users’ group affiliations on one political topic predict their affiliations on another topic, capturing how stances align across issues. We measure alignment using normalized mutual information (NMI), which ranges from 0 (no alignment) to 1 (perfect alignment). Higher NMI values indicate that knowing a user’s group on one issue provides substantial information about their group on another, reflecting strong cross-issue alignment.  \\
\end{metricbox}

\section{Findings}
\label{sec:findings}

We begin by presenting a brief overview of descriptive statistics on Bluesky activitity patterns, followed by an analysis of the platform’s political content. We then report our results on polarization.

\subsection{Overview of Bluesky}

\paragraph{Activity patterns}

During our data collection, we recorded the creation of 13.5M new profiles on the platform, an average of 82K sign-ups per day, which demonstrates the platform’s increased popularity. User engagement is already substantial: We logged over 1B new posts, nearly 1B reposts, and over 6B likes. All statistics are detailed in Table \ref{tab:bluesky_activity}.

\begin{table}[htbp]
\centering
\caption{\textbf{Basic Bluesky statistics.} Daily activity and total counts for \textit{create} actions as recorded during our data collection, grouped by action type.}
\renewcommand{\arraystretch}{1}
\begin{tabular}{lcc@{\hspace{1cm}}cc}
\toprule
\multirow{2}{*}{\textbf{Action Type}} & \multicolumn{2}{c}{\textbf{Daily Average}} & \multicolumn{2}{c}{\textbf{Total Count}} \\
\cmidrule(lr){2-3}\cmidrule(lr){4-5}
& \textbf{Actions} & \textbf{Authors} & \textbf{Actions} & \textbf{Authors} \\
\midrule
Likes & 36\,783\,658 & 1\,867\,068 & 6\,032\,519\,995 & 306\,199\,225 \\
Posts & 6\,227\,244 & 1\,005\,727 & 1\,021\,268\,020 & 164\,939\,388 \\
Reposts & 5\,284\,612 & 649\,805 & 866\,676\,384 & 106\,568\,078 \\
Blocks & — & 154\,927 & — & 25\,408\,088 \\
Follows & — & 946\,840 & — & 155\,281\,833 \\
Sign-ups & — & 82\,319 & — & 13\,500\,391 \\
\bottomrule
\end{tabular}
\label{tab:bluesky_activity}
\end{table}

As illustrated in Figs. \ref{fig:posts_activities_per_day} and~\ref{fig:users_activities_per_day} in the Appendix, daily activity levels remain steady, with the notable exception of follow actions. Follows surged until early 2025, likely spurred by several events: Bluesky lifting its invite-only restriction, the introduction of Starter Packs in June 2024, the ban of Twitter/X in Brazil (August–October 2024), and a wave of users departing Twitter/X after the November 2024 U.S. election \cite{quelle2025academicsleavingtwitterbluesky, hinsliff2024exodus}.

%\begin{table}[h]
%    \centering
%    \caption{Average daily Bluesky activity  of accounts, as recorded during our data collection per action type for \textit{create} actions of the selected types included in the analysis.}
%    \begin{tabular}{ c c }
%        \hline
%        \textbf{Action Type} & \textbf{Average Number of Authors per Day} \\
%        \hline
%        Likes & 1\,867\,068\\ 
%        Posts & 1\,005\,727\\
%        Reposts & 649\,805\\
%        Blocks & 154\,927\\
%        Follows & 946\,840\\
%        Sign-ups & 82\,319\\
%        \hline
%        \end{tabular}
%        \label{tab:avg_daily_authors_stats}
%    \end{table}

%\begin{table}[h]
%    \centering
%   \caption{Average daily Bluesky activity  related to posts, as recorded during our data collection per action type for \textit{create} actions of the selected types included in the analysis.}
%   \begin{tabular}{ c c }
%       \hline
%       \textbf{Action Type} & \textbf{Average Number of Actions per Day} \\
%       \hline
%       Likes & 36\,783\,658\\
%       Posts & 6\,227\,244\\
%       Reposts & 5\,284\,612\\
%       \hline
%       \end{tabular}
%       
%       \label{tab:avg_daily_posts_stats}
%   \end{table}

\paragraph{Topic landscape}

We begin by exploring the range of topics discussed on Bluesky as illustrated in Fig. \ref{fig:pipeline_schema} and described in Section \ref{secsec:finding-topics}. Our filtered dataset contains over 43 million posts, of which 3\% (about 1 290 000 posts) were used to identify semantically coherent subtopics. This process produced 81 distinct subtopics. The largest non-political subtopics in terms of the size of these semantic clusters included \textit{Sexual Content \& Fetishes}, \textit{Game Development Trends}, and \textit{Bluesky Community Networking}, while the most prominent political subtopics were \textit{Ukraine-Russia Conflict}, \textit{US-Canada Trade Tensions}, and \textit{Democratic Party Frustration (US)}. Some examples of relatively niche topics were e.g. \textit{ABDL Lifestyle} (Adult Baby/Diaper Lover community), \textit{Furry Fursuit Community} (Furry fandom and fursuit community content) and \textit{Distorted Christianity} (Critique of misuse of Christian beliefs). A complete list of identified subtopics is provided in Appendix \ref{appendix-subtopics}.

%It is important to note that the size of a semantic cluster does not necessarily reflect the real-world prevalence of a topic; rather, large clusters may result when many posts share similar semantic features. Nonetheless, semantic clustering remains a standard and efficient method for mapping the diversity of topics from any textual dataset. 

\subsection{Politics on Bluesky}

\begin{table}[ht]
\centering
\caption{\textbf{Distribution of posts by political theme.} Each post in the filtered dataset was classified into one of the following categories, adapted from the \textit{Comparative Agendas Project} (CAP) Codebook \cite{jones2023policy}.}
\begin{tabular}{@{}lccc@{}}
\toprule
\textbf{Category}              & \textbf{\% of All Posts} & \textbf{\% of Political Posts} & \textbf{Number of Posts} \\
\midrule
Apolitical                    & 87.3\%   & ---     & 38 129 599   \\
Political                     & 12.7\%   & 100\%   & 5 522 980   \\
\quad Civil Rights      & 3.5\%   & 27.3\%  & 1 509 130   \\
\quad Defense \& International Affairs     & 3.6\%   & 28.3\%  & 1 560 553   \\
\quad Economy, Trade \& Labor               & 1.6\%   & 12.6\%  & 693 207   \\
\quad Government Operations \& Administration   & 0.8\%   & 6.0\%  & 332 586   \\
\quad Infrastructure \& Environment               & 0.5\%   & 3.8\%  & 209 171   \\
\quad Law, Crime \& Justice          & 2.0\%   & 16.1\%  & 891 571   \\
\quad Science, Technology \& Energy   & 0.3\%   & 2.4\%  & 131 114   \\
\quad Social Policy          & 0.4\%   & 3.5\%  &  195 648   \\
\midrule
Total                         & 100\%  & ---     & 43 652 579 \\
\bottomrule
\end{tabular}
\label{tab:post_distribution}
\end{table}

Approximately 13\% (5 522 980) of posts in our filtered dataset were classified as political. The most prevalent political themes were \textit{Defense \& International Affairs}, \textit{Civil Rights}, and \textit{Law, Crime \& Justice} (see Table \ref{tab:post_distribution}). In contrast, the least represented political themes were related to \textit{Science, Technology \& Energy}, and \textit{Social Policy}.

To select the parent topics for further analysis, we filtered the subtopics based on their political content. Of the 81 subtopics, 35 contained more than three-quarters of posts labeled as non-political and were therefore classified as apolitical. The remaining 46 subtopics, which had a higher proportion of political content, were classified as political (see Appendix Table \ref{appendix-subtopics}). Finally, based on these 46 political subtopics we selected ten parent topics that we believe represent the majority of these subtopics: \textit{Trump administration}, \textit{Elon Musk}, \textit{US-Canada relations}, \textit{LA wildfires}, \textit{DEI programs}, \textit{TikTok ban}, \textit{Israel–Palestine}, \textit{Russia–Ukraine}, \textit{LGBTQ+ rights}, and \textit{AI}. 

These topics were selected for their current prevalence and relevance, and were grouped into three broad categories: U.S.-related timely issues, ongoing international conflicts, and socio-technological issues. We believe this selection of topics together provide a comprehensive overview of the major themes present both in the dataset and today's world.  Networks for these topics were constructed as described in Section \ref{network_construction}, and their basic properties are summarized in Table \ref{tab:network_properties}.

\begin{table}[ht]
\centering
\caption{\textbf{Network properties of the ten selected political topics.} The table reports the number of nodes, number of edges, and average degree for each topic-specific discourse network.}
\begin{tabular}{@{}lccc@{}}
\toprule
\textbf{Topic} & \textbf{Number of Nodes (|$V^{\text{reposts}}$|} & \textbf{Number of Edges (|$E^{\text{reposts}}$|)} & \textbf{Average Degree} \\
\midrule
Trump administration         & 990 893  & 30 705 827  & 61.98   \\
Elon Musk                    & 485 135  & 4 654 921  & 19.19   \\
US-Canada relations          & 208 884  & 1 078 015  & 10.32   \\
LA wildfires                 & 165 078  & 593 857  & 7.19   \\
DEI programs                 & 284 571  & 1 039 771  & 7.31   \\
TikTok ban                   & 134 614  & 255 452  & 3.80   \\
\midrule
Israel--Palestine            & 276 322  & 1 660 527  & 12.02   \\
Russia--Ukraine              & 390 198  & 4 611 576  & 23.64   \\
\midrule
LGBTQ+ rights                & 375 677  & 1 709 459  & 9.10   \\
AI                           & 169 919  & 407 215  & 4.79   \\
\bottomrule
\end{tabular}
\label{tab:network_properties}
\end{table}

Among the constructed networks, the U.S.-related networks, \textit{Trump administration} and \textit{Elon Musk}, are the largest, with the former containing approximately one million nodes and the latter about half a million. In contrast, the smallest networks are \textit{TikTok ban}, \textit{LA wildfires}, and \textit{AI}, each comprising fewer than 200 000 nodes.

\begin{figure}[t!]
    \centering
    \begin{subfigure}[t]{0.49\textwidth}
        \centering
        \includegraphics[width=\textwidth]{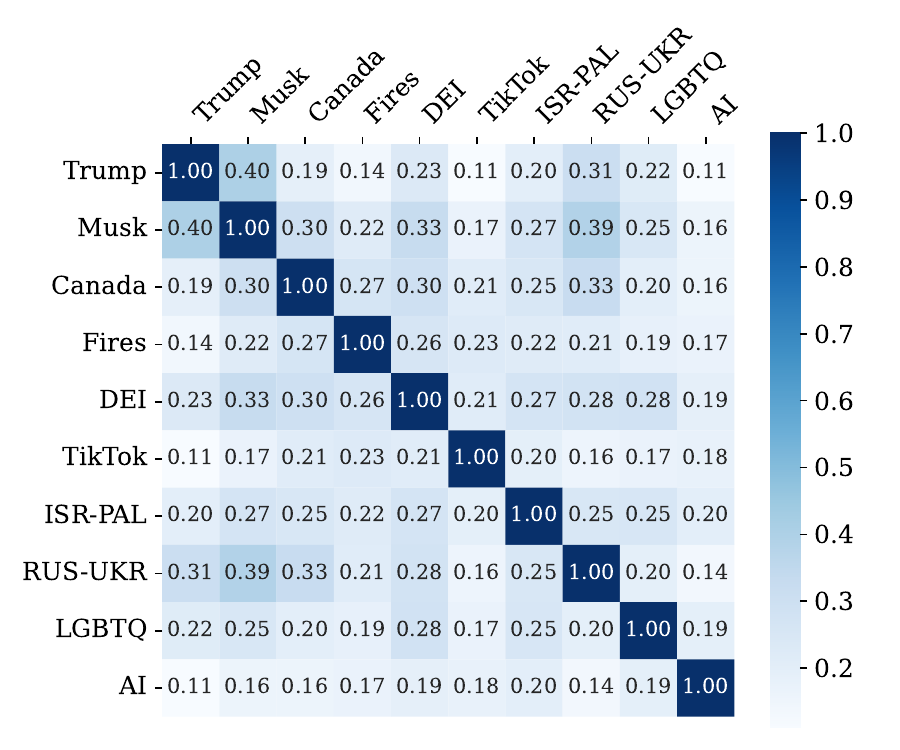}
        \caption{Heatmap of user overlap between topic pairs, quantified by the Jaccard index. Higher values indicate greater overlap in user participation between the respective topics.}
        \label{fig:subplot1}
    \end{subfigure}
    \hfill
    \begin{subfigure}[t]{0.49\textwidth}
        \centering
        \includegraphics[width=\textwidth]{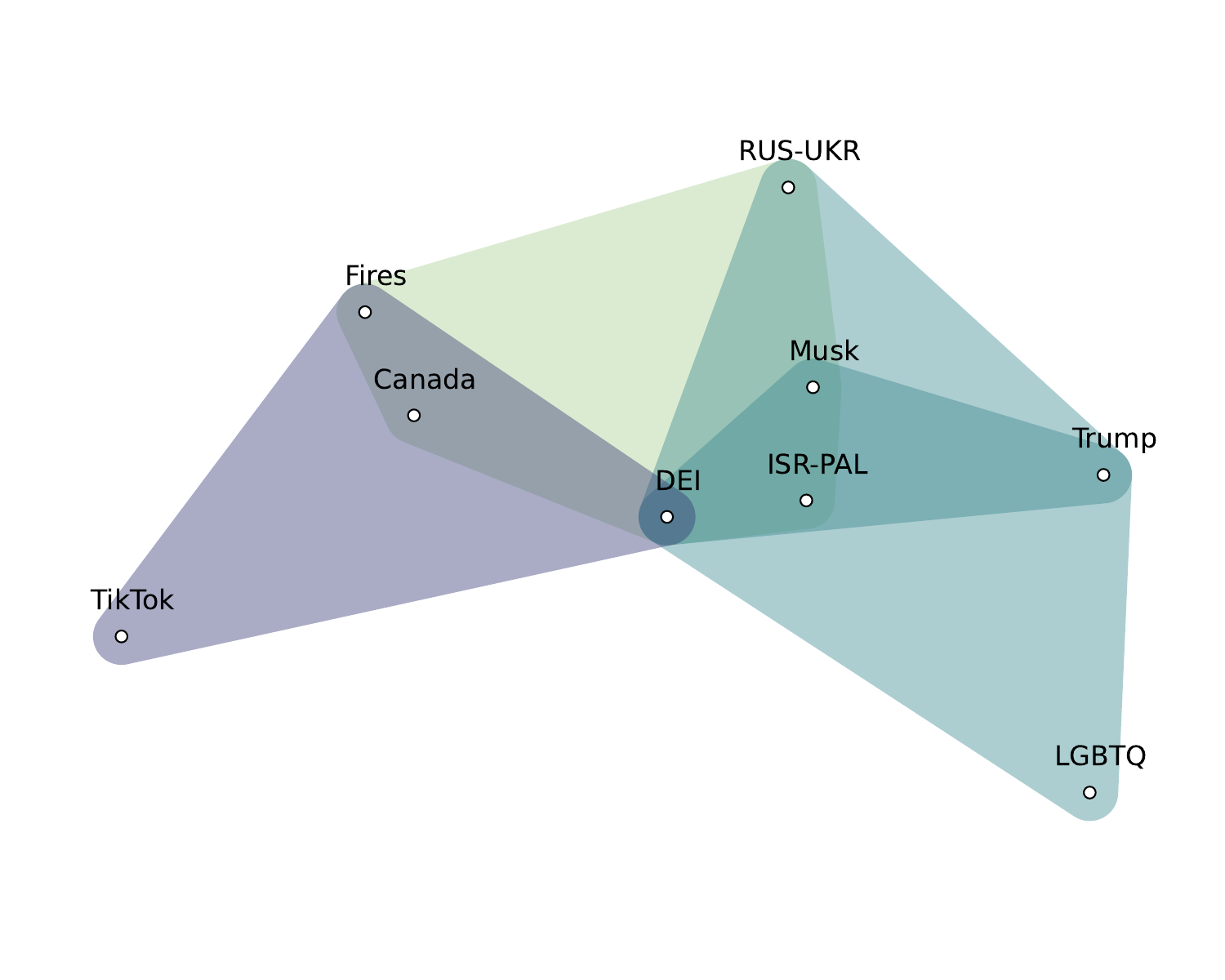}
        \caption{Hypergraph representation of topic bundles based on user overlap. Each node represents a topic, and an edge is drawn between two or more topics if the Jaccard index for user overlap between them exceeds a specified threshold of 0.2. Connected topics indicate a shared user base actively participating in discussions on topics.}
        \label{fig:subplot2}
    \end{subfigure}
    \caption{\textbf{User overlap between political topics on Bluesky.}}
    \label{fig:mainfigure-useroverlap}
\end{figure}

To understand how these political topics are interconnected, we analyzed user overlap between pairs of topics (see Fig. \ref{fig:mainfigure-useroverlap}a). We found high overlap in user participation among the \textit{Trump administration}, \textit{Elon Musk}, and \textit{Russia-Ukraine} networks, with pairwise overlaps exceeding 0.30 (Jaccard indxex) in all cases. This indicates that users often engage with multiple topics in these domains simultaneously. In contrast, the user bases for \textit{AI} and \textit{TikTok ban} showed the lowest overlaps with other topics, with the interesting exception of \textit{TikTok ban} and \textit{LA wildfires}, which had a Jaccard index of 0.23.

The hypergraph representation reveals several topic bundles (see Fig. \ref{fig:mainfigure-useroverlap}b). A hyperedge was added between topics if they formed a clique in which all pairwise overlaps (dyad weights) exceeded 0.2. This resulted in four distinct hyperedges: \{\textit{DEI}, \textit{ISR-PAL}, \textit{Musk}, \textit{Trump}, \textit{LGBTQ}\}, \{\textit{DEI}, \textit{ISR-PAL}, \textit{Musk}, \textit{Trump}, \textit{RUS-UKR}\}, \{\textit{DEI}, \textit{ISR-PAL}, \textit{Musk}, \textit{RUS-UKR}, \textit{Fires}, \textit{Canada}\} and \{\textit{DEI}, \textit{TikTok}, \textit{Canada}, \textit{Fires}\}

\textit{DEI}, \textit{ISR-PAL}, and \textit{Musk} appear in almost every bundle, indicating these topics have substantial user overlap and are central to the network of political conversations. The bundle of \{\textit{DEI}, \textit{TikTok}, \textit{Canada}, \textit{Fires}\} is somewhat distinct from the larger, more geopolitically oriented clusters. While these topics are still highly political, they may represent a different dimension of political discussion -- one that centers more on cultural issues and policy debates, rather than partisan conflict or international affairs. 

\begin{figure}
    \centering
    \begin{overpic}[width=\textwidth]{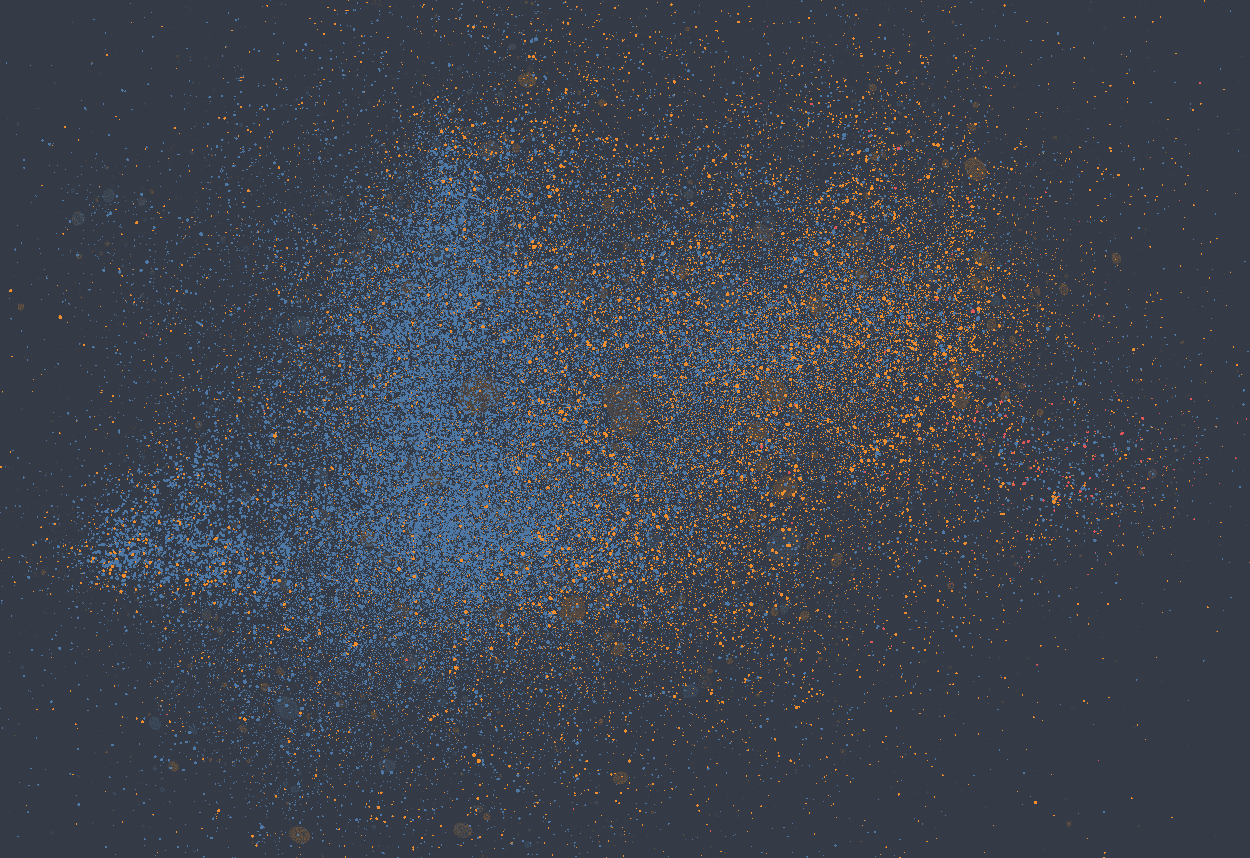}
        % Place the overlay image at (x,y) coordinates (in % of width and height)
        %\put(0,41){\includegraphics[width=0.4\textwidth]{figures/network_graph_isrpal_structural.png}}
    \put(1.2,40){\setlength{\fboxrule}{0.1pt}\setlength{\fboxsep}{0pt}\fcolorbox{white}{white}{\includegraphics[width=0.4\textwidth]{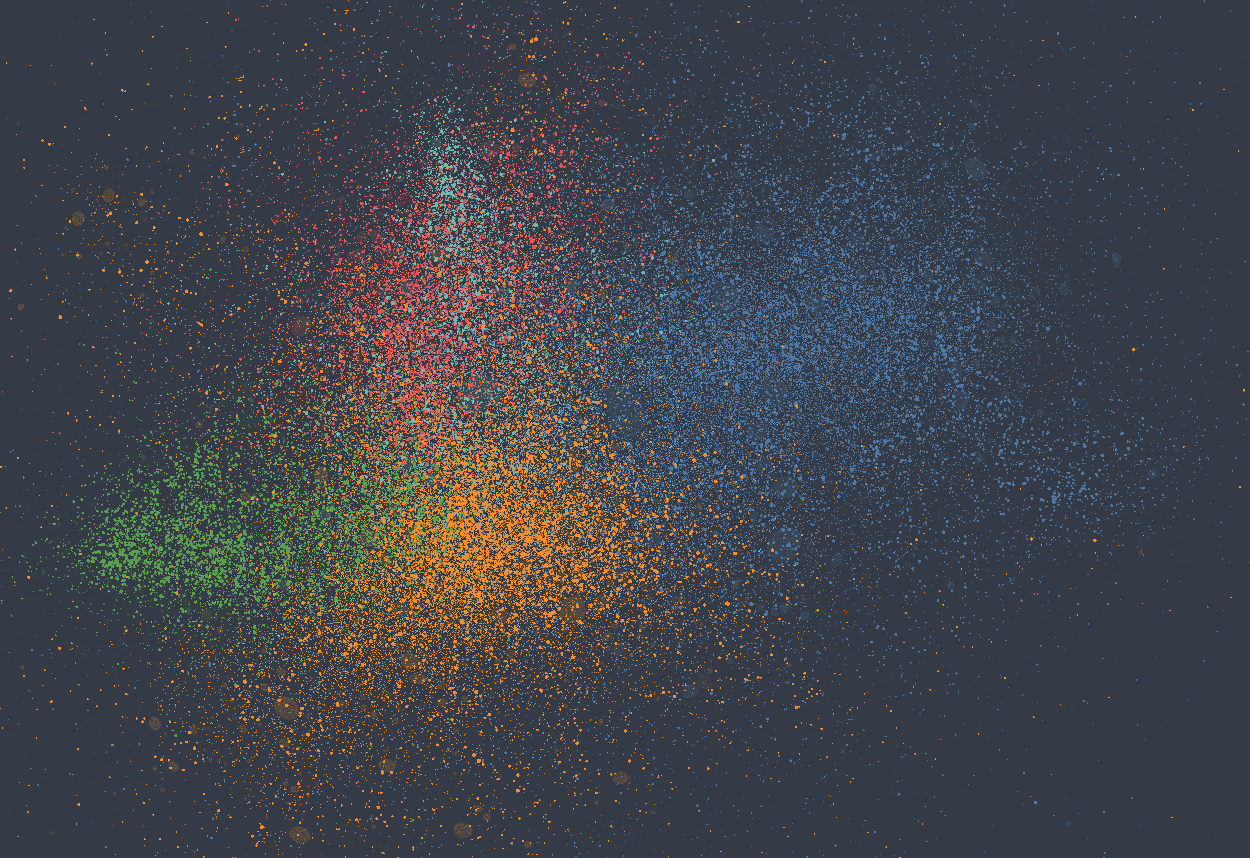}}}        
    \end{overpic}
    \caption{\textbf{Repost network of Israel-Palestine.} In the main plot, nodes are colored according to stance annotations inferred using the content-based approach described in Section \ref{content-based-detection} (blue: pro-Palestine, orange: neutral, red: pro-Israel). The inset in the upper left shows the same network, with nodes colored by structural groups detected via a stochastic block model (see Section \ref{structural-based-detection}). For clarity, nodes with only one connection are omitted from both visualizations. Repost network of Russia-Ukraine can be found in Appendix \ref{appendix-rusukr-graph}. Networks are visualized in \textit{Cosmograph} with default colors.}
    \label{isrpal-graph-viz}
\end{figure}

\subsection{Polarization on Bluesky}

Having explored activity patterns, mapped the landscape of topics and political themes, and uncovered connections across different domains on Bluesky, we aim to address another central question: how homogeneous or polarized are these conversations in terms of political stances? Can we identify groups with opposing stances, and if so, how balanced are they, for instance, in terms of size?

\subsubsection{Polarization patterns within networks}

We begin by computing, for each network, the same set of scores using groups inferred via the content-based approach. In this method, groups are defined solely by the content a node has created or engaged with (see \ref{group-detection}). The LLM (Mistral 24B) is prompted to classify each user’s stance as either \textit{for}, \textit{neutral}, or \textit{against}, resulting in a maximum of three distinct, non-overlapping groups per topic. These labels are tailored to each topic's context. For example, for the \textit{Russia-Ukraine} discussion, the labels are \textit{supports\_ukraine}, \textit{supports\_russia}, and \textit{neutral}; for the \textit{Trump administration} network, the labels are \textit{supports\_trump}, \textit{opposes\_trump}, and \textit{neutral}. Measurements for each network are reported in Table \ref{tab:polarization_scores}.

\begin{table}[ht]
\centering
\caption{
\textbf{Polarization and diversity scores for major topics discussed on Bluesky.}
Topics are grouped as follows: U.S.-related issues (top section), ongoing international conflicts (middle), and other social/technological topics (bottom). 
Columns show the fraction of LLM-annotated stance groups (A, Neutral, B), Simpson diversity index, assortativity, AEI polarization, and Coleman index for both A and B (see Methods for metric definitions). 
A indicates the majority stance group for each topic, while B is the minority group (see Methods for details on group assignment).
}

\begin{tabular}{lccccccccc}
\toprule
Topic & $\%$ A & $\%$ Neutral & $\%$ B & Simpson & Assort. & AEI & Col. A & Col. B & Dominant Stance \\
\midrule
% --- US-related topics ---
Trump admin           & 0.82 & 0.17 & 0.01 & 0.02 & 0.08 & \textbf{0.74} & 0.82 & 0.07 & Opposes Trump \\
Elon Musk             & 0.80 & 0.20 & 0.01 & 0.01 & 0.10 & \textbf{0.88} & 0.83 & 0.09 & Opposes Musk \\
US-Canada             & 0.47 & 0.46 & 0.07 & 0.23 & 0.10 & 0.43 & 0.56 & 0.04 & Supports Canada \\
LA wildfires          & 0.19 & 0.77 & 0.04 & 0.29 & 0.20 & 0.31 & 0.13 & 0.03 & Neutral \\
DEI programs          & 0.53 & 0.27 & 0.20 & 0.40 & 0.12 & 0.30 & 0.55 & 0.02 & Supports DEI \\
TikTok ban            & 0.24 & 0.67 & 0.08 & 0.38 & 0.54 & 0.62 & 0.25 & 0.31 & Neutral \\
\midrule
% --- Wars/conflicts ---
Israel–Palestine      & 0.59 & 0.41 & 0.00 & 0.01 & 0.30 & \textbf{0.95} & 0.71 & 0.24 & Supports Palestine \\
Russia–Ukraine        & 0.71 & 0.29 & 0.01 & 0.02 & 0.22 & \textbf{0.90} & 0.76 & 0.17 & Supports Ukraine\\
\midrule
% --- Other topics ---
LGBTQ+ rights         & 0.75 & 0.20 & 0.05 & 0.11 & 0.05 & 0.45 & 0.77 & 0.01 & Supports LGBTQ+ \\
AI                    & 0.27 & 0.72 & 0.02 & 0.12 & 0.26 & 0.59 & 0.40 & 0.02 & Neutral \\
\bottomrule
\end{tabular}
\label{tab:polarization_scores}
\end{table}

The highest polarization is observed in networks focused on international conflicts: both \textit{Israel–Palestine} (AEI = 0.95) and \textit{Russia–Ukraine} (AEI = 0.90) have the highest AEI polarization scores in the table. These findings align with previous observations of heightened polarization around the conflicts on Bluesky \cite{quelle2025bluesky}. These are followed by the \textit{Elon Musk} (AEI = 0.88) and \textit{Trump administration} (AEI = 0.74) networks, which also exhibit substantial polarization. These topics also have the lowest diversity (Simpson index of 0.02 or less), indicating that nearly all posts are concentrated in a single stance group. This can also be seen directly from the \% B column, which shows that the minority stance group constitutes 1\% or less in all cases. In other words, the dominant stances in these networks, such as supporting Palestine or Ukraine, or opposing Musk or Trump, are significantly larger than the groups holding opposing views, and the two sides are strongly isolated from each other, with limited engagement or overlap between stance groups.

In contrast to the highly polarized topics, discussions around \textit{US–Canada} relations, \textit{DEI programs} and \textit{LGBTQ+ rights} exhibit a slightly more balanced distribution of stances. These topics also have higher Simpson diversity and lower AEI polarization scores compared to topics dominated by a single stance, suggesting more balanced and diverse conversations. Conversely, topics such as \textit{LA wildfires}, \textit{TikTok ban}, and \textit{AI} are marked by a large proportion of neutral stances (77\%, 67\%, and 72\%, respectively), with no strong majority group emerging. Again, the prevalence of higher Simpson diversity and lower AEI in these cases indicates that discussions are more mixed and less polarized, possibly reflecting a broader range of perspectives and a lack of clear division between opposing camps.

A similar analysis was performed for the purely structure-based groups, with the main scores reported in Table \ref{tab:structural_pol_table}. The most notable observation is that, particularly for the most polarized topics, multiple subcommunities with the same stance were identified. These subcommunities can also be highly isolated from each other, as illustrated in Table \ref{tab:structural_pol_table} and Fig. \ref{isrpal-graph-viz}. We also report the AEI scores between every structural groups identified in each network in Appendix \ref{appendix-aei-matrix}.

\begin{table}[ht]
\centering
\caption{\textbf{Polarization and other group structure related scores for major topics discussed on Bluesky.} In this table, the groups were inferred purely based on structure via stochastic block model. As number of groups in the system can vary between 1-5, the mean, max and minimum value observed between groups are reported. \% DS is the fraction of users with the globally dominant stance in a structural group.}
\label{tab:structural_polarization_scores}
\begin{tabular}{lcccccc}
\toprule
Topic & Mean AEI & Max. AEI & Min. AEI & Number of Groups & Max. \% DS & Min. \% DS \\
\midrule
Trump admin      & 0.70 & 0.79 & 0.63 & 3 & 0.99 & 0.78 \\
Elon Musk        & 0.75 & 0.87 & 0.58 & 4 & 0.99 & 0.76  \\
US-Canada        & 0.73 & 0.86 & 0.48 & 3 & 0.93 & 0.44 \\
LA wildfires     & 0.80 & 0.82 & 0.76 & 3 & 0.82 & 0.73  \\
DEI programs     & 0.78 & 0.81 & 0.74 & 3 & 0.83 & 0.52 \\
TikTok ban       & -- & -- & -- & 1 & -- & --  \\
\midrule
Israel--Palestine & 0.84 & 0.97 & 0.65 & 5 & 0.80 & 0.48  \\
Russia--Ukraine   & 0.77 & 0.97 & 0.55 & 5 & 0.78 & 0.65 \\
\midrule
LGBTQ+ rights     & 0.81 & 0.91 & 0.67 & 5 & 0.99 & 0.69 \\
AI                & 0.94 & 0.94 & 0.94 & 2 & 0.93 & 0.70  \\
\bottomrule
\end{tabular}
\label{tab:structural_pol_table}
\end{table}

\subsubsection{Polarization patterns between networks}

\begin{figure}[t]
    \centering
    \begin{subfigure}[t]{0.49\textwidth}
        \centering
        \includegraphics[width=\textwidth]{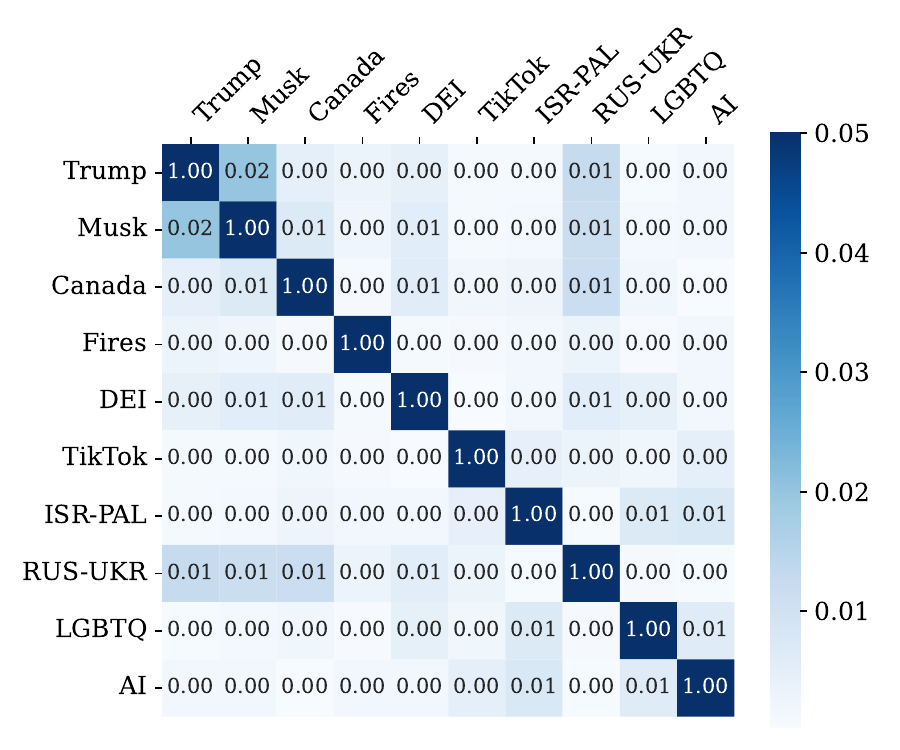}
        \caption{Issue alignment for content-based groups}
        \label{fig:subplot1}
    \end{subfigure}
    \hfill
    \begin{subfigure}[t]{0.49\textwidth}
        \centering
        \includegraphics[width=\textwidth]{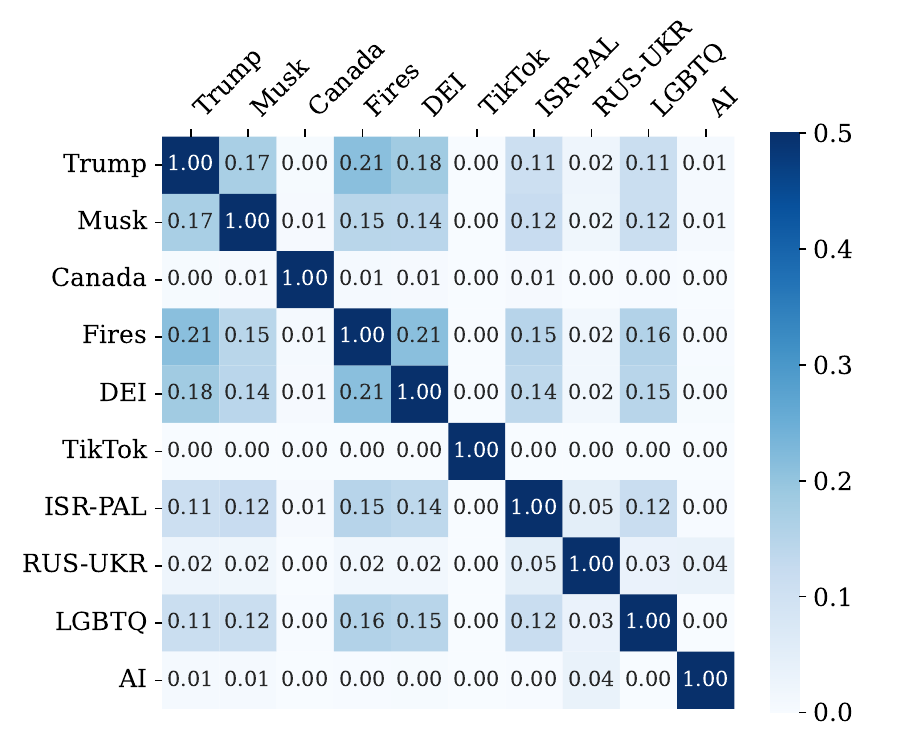}
        \caption{Issue alignment for structure-based groups}

        \label{fig:subplot2}
    \end{subfigure}
    \caption{\textbf{Issue alignment across topic pairs via normalized mutual information.} Heatmaps showing issue alignment between pairs of topics, quantified by \textit{normalized mutual information} (see Section \ref{political-polarization-section}). The left heatmap, based on content-inferred groups, displays generally low alignment values. In contrast, the right heatmap shows results for groups inferred structurally, which allows for subcommunities within larger communities that share similar views. Slightly higher alignment across networks is observed in this case.}
    \label{fig:both_alignment}
\end{figure}

We find that issue alignment between stance groups is generally low across topics, particularly when groups are inferred from content (see Fig. \ref{fig:both_alignment}). The left heatmap shows that, for most topic pairs, there is little or no consistency in users’ stances across issues, suggesting that individuals’ positions tend to be topic-specific. In contrast, the structure-based grouping displayed in the right heatmap shows slightly higher, though still modest, alignment across some topics. For example, there is more overlap between \textit{Trump} and \textit{Musk}, or \textit{DEI} and \textit{LGBTQ+} rights.

This suggests that structural communities can capture clusters of users whose views or attitudes appear more consistently aligned across issues, reflecting broader subcommunities within the network. However, it is important to note that these structural groups may not necessarily represent opposing viewpoints; rather, they may be groups of users who share similar overall attitudes but are clustered differently based on their interaction patterns. Whether this is driven by algorithmic factors or by genuine differences in the nuances or perspectives from which groups approach topics remains unclear. Nevertheless, the overall low levels of normalized mutual information indicate that strong issue polarization remains largely confined to individual topics, with only limited evidence of cross-topic alignment.

\begin{figure}[h]
    \centering
    \begin{subfigure}[t]{0.49\textwidth}
        \centering
        \includegraphics[width=\textwidth]{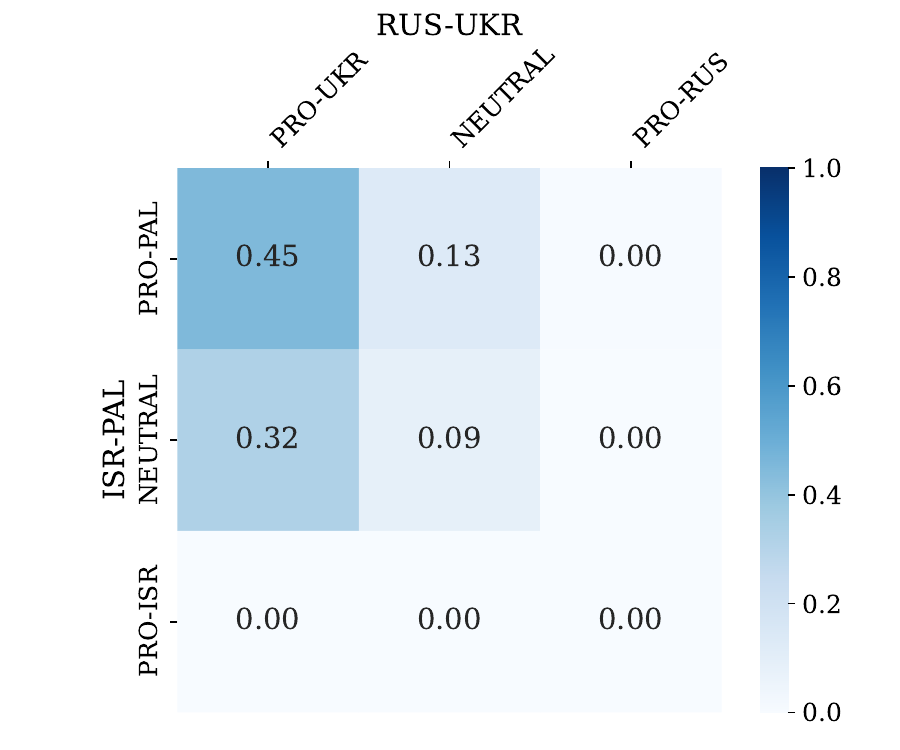}
        \caption{Russia-Ukraine \& Israel-Palestine}
        \label{fig:subplot1}
    \end{subfigure}
    \hfill
    \begin{subfigure}[t]{0.49\textwidth}
        \centering
        \includegraphics[width=\textwidth]{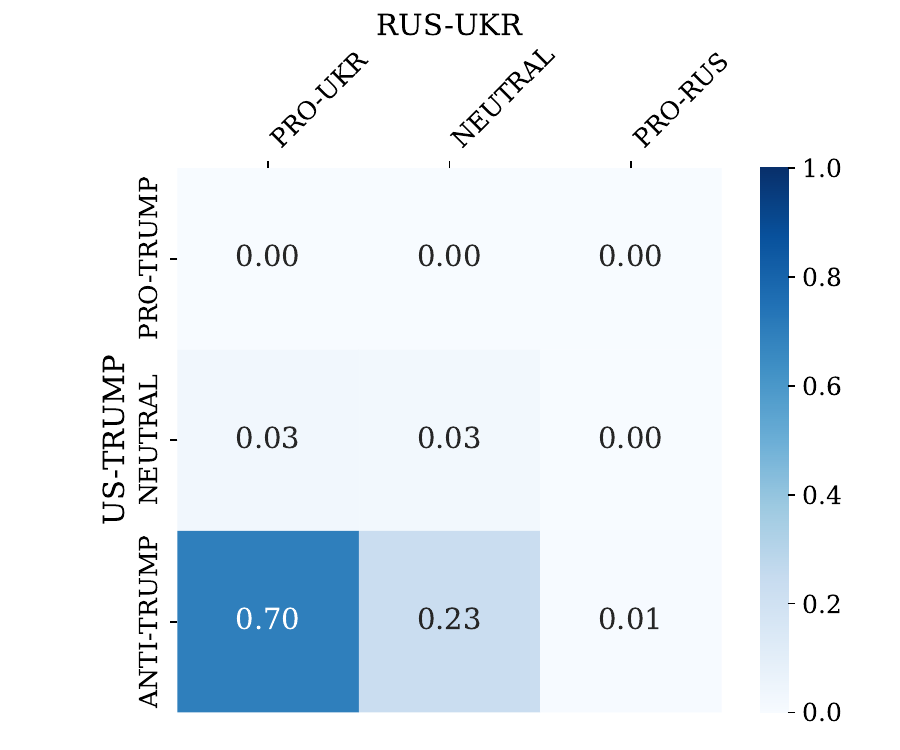}
        \caption{Russia-Ukraine \& US-Trump}
        \label{fig:subplot2}
    \end{subfigure}
    \caption{\textbf{Joint probability table for joint political stances}. Each cell shows the proportion of individuals taking the corresponding combination of stances on both given topics. Find the tables for the remaining pairs in Appendix \ref{appendix-joint-probs-tables}.}
    \label{fig:jointprobsmain}
\end{figure}

We also report joint probability tables (see Table \ref{fig:jointprobsmain}) for the four most polarized networks to provide a direct view of the overlap between stance groups. While these joint probabilities may primarily reflect group size effects rather than genuine alignment, they help illustrate concrete patterns of co-occurrence of stances. NMI remains the more rigorous and interpretable measure of alignment, particularly when controlling for randomness or group size imbalance.

We observe that, while the most probable stance pairs reflect the dominant positions within each network, the pattern of neutrality is not symmetrical across topics. For example, as shown in Fig. \ref{fig:jointprobsmain}a, an individual is more than twice as likely to hold a pro-Ukraine stance while remaining neutral on Israel–Palestine than to hold a pro-Palestine stance while being neutral on Russia–Ukraine. Similar asymmetries are observed across all topic pairs (see Figs. \ref{fig:jointprobsmain}b, \ref{fig:appendix-joint-probs1}, and \ref{fig:appendix-joint-probs2}).

\begin{figure}[htbp]
    \centering
    \begin{subfigure}[t]{0.49\textwidth}
        \centering
        \includegraphics[width=\textwidth]{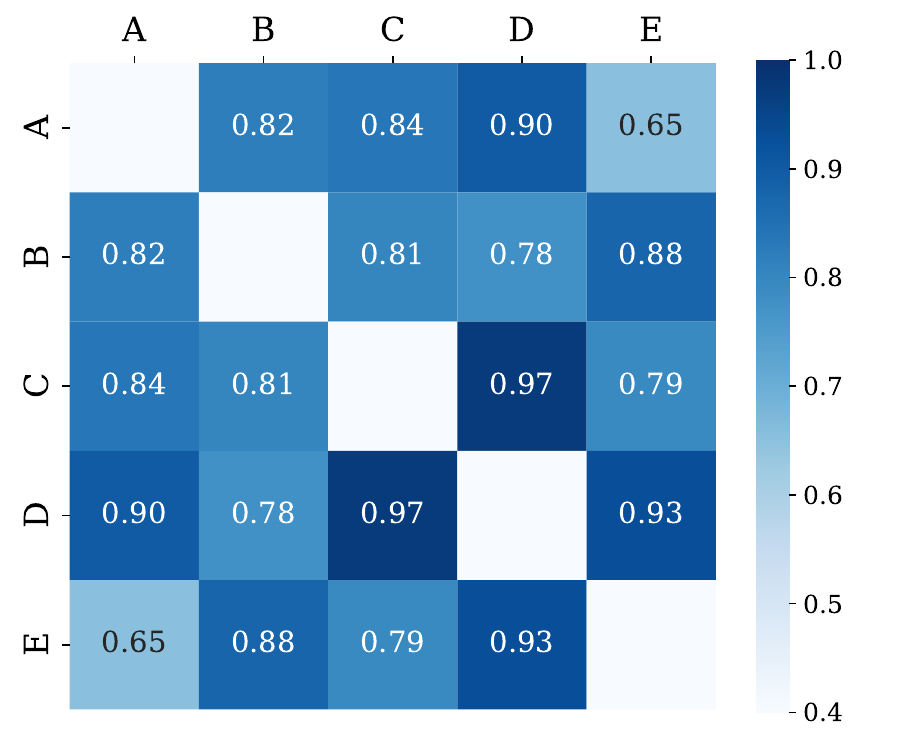}
        \caption{Israel-Palestine network}
        \label{fig:subplot1}
    \end{subfigure}
    \hfill
    \begin{subfigure}[t]{0.49\textwidth}
        \centering
        \includegraphics[width=\textwidth]{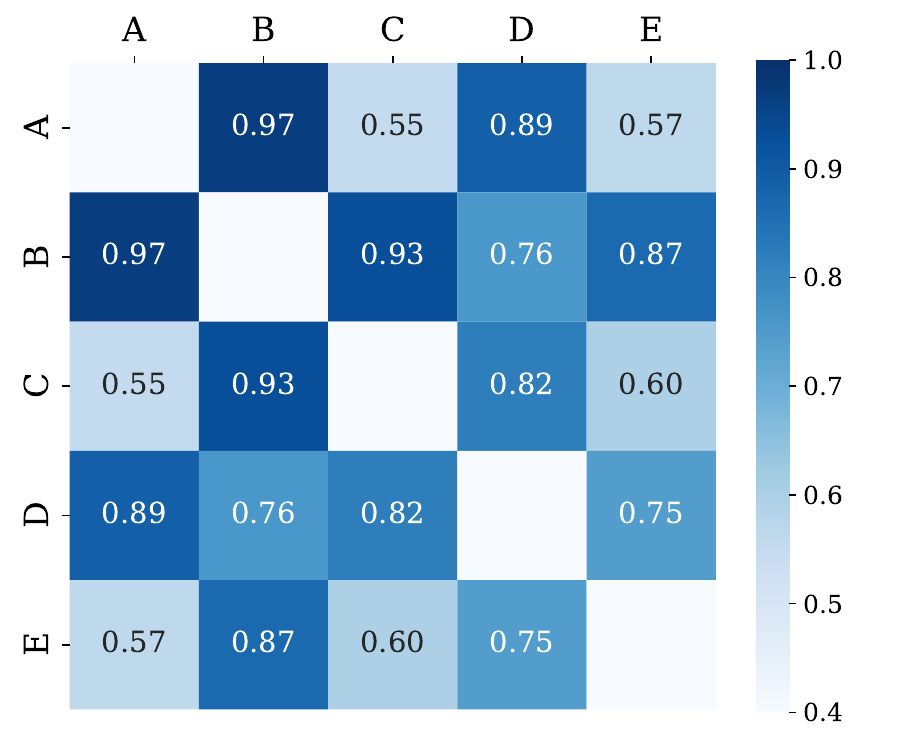}
        \caption{Russia-Ukraine network}
        \label{fig:subplot2}
    \end{subfigure}
    \caption{\textbf{AEI polarization scores between structural groups.} Heatmap of AEI polarization scores between structural groups (labeled A--E) for the networks representing communication patterns on ongoing international conflicts. Higher values indicate greater division between group pairs. Find the heatmaps for the remaining topics in Appendix \ref{appendix-aei-matrix}.}
    \label{fig:mainfigure}
\end{figure}

\section{Conclusion}
\label{sec:conclusion}

In this study, we set out to systematically map and analyze the landscape of political discourse and polarization on Bluesky. Our goal was to identify what kinds of topics, both political and non-political, are being discussed, and to rigorously assess whether patterns of polarization exist on the platform. 

Our findings show that Bluesky hosts a wide variety of topics, spanning both political and non-political domains. Non-political conversations are especially rich and diverse, with many topics revolving around hobbies, arts, gaming, and community networking. After systematically classifying and filtering the data, we focused our analysis on ten prominent political topics, each highly relevant to current affairs, mostly related to international conflicts, U.S. politics, and civil rights. For these, we constructed network representations to uncover patterns of interaction and polarization.

We found clear evidence of polarization in discussions around certain timely issues, most notably international conflicts (\textit{Russia–Ukraine} and \textit{Israel–Palestine}), as well as highly visible topics like the \textit{Trump Administration} and \textit{Elon Musk}. However, while strong polarization exists, the minority stance is consistently less prevalent than the majority stance, often by considerable margins.

In contrast, some topical issues such as \textit{AI} and the \textit{TikTok} ban showed notably low levels of polarization, with large proportions of users holding a neutral stance and limited evidence of deep divisions. Across the platform more broadly, we found little indication of strong issue alignment between user groups: Individuals’ stances tended to be topic-specific, rather than consistently aligned across multiple issues. That said, we did observe some alignment among structurally defined groups, implying that certain communities may generally share similar attitudes, though possibly with varying degrees of criticality or emphasis. Whether this reflects Bluesky’s algorithmic design or organic user dynamics is an open question and presents an important direction for future research.

A rigorous, large-scale validation of LLM-assigned stances is beyond the scope of this study, but remains a crucial goal for future work. Here, we relied on targeted spot-checks for each major topic, which showed that precision and recall were generally reasonable for our initial analysis. However, thorough validation with a formal codebook and human annotation will be necessary to fully confirm reliability. Misclassifications most often involved posts featuring sarcasm or very short texts---a well-known challenge for LLMs. We also found that posts containing simple headlines or event statements were frequently labeled as neutral, which may or may not be desirable depending on one's definition of neutrality. Classification performance improved when we aggregated multiple posts per user (or their network neighbors) at the node level, rather than classifying posts individually. 

Overall, Bluesky is emerging as a rapidly growing venue for substantive political communication. While some polarization patterns familiar from other social media platforms are already observable, our results strongly suggest that the user base is more homogeneous, with single stance groups often dominating discussions. Whether this will change over time, potentially converging toward the polarization patterns seen on established platforms such as Twitter or Facebook, remains to be seen.

One possible interpretation of our findings is that we are witnessing not just polarization within individual platforms, but the culmination of a broader, ecosystem-wide polarization across the digital landscape. Rather than encountering deeply divided groups on a single platform, we may be seeing the emergence of distinct online spaces, such as Truth Social or, increasingly, Twitter/X after recent trends in demographic shifts, where users largely agree on political issues. In this scenario, digital platforms themselves are becoming more politically homogeneous, each catering to relatively like-minded communities. As a result, the clash of conflicting views is diminishing, not because the surrounding polarization has decreased, but because individuals with differing political attitudes are increasingly segregated into separate, self-reinforcing digital environments. 

\section*{Acknowledgments}
The calculations presented above were performed using computer resources
within the Aalto University School of Science ``Science-IT'' project. We would also like to thank Corinna Coupette for their valuable assistance in proofreading and providing helpful comments on earlier drafts of this manuscript.

\bibliographystyle{plain}
\bibliography{references}  
\clearpage
\appendix

\setcounter{table}{0}
\renewcommand\thetable{\thesection.\arabic{table}}

\setcounter{figure}{0}
\renewcommand\thefigure{\thesection.\arabic{figure}}

%\section*{Appendix}
\part*{Appendix}

\section{Political themes}
% --- TABLE:    Political thems and their descriptions ---
\begin{table}[h!]
\centering
\caption{High-level political themes for classifying posts, adapted from the CAP Codebook \cite{jones2023policy}.}
\begin{tabular}{@{}lp{10cm}@{}}
\toprule
\textbf{Category} & \textbf{Description} \\
\midrule
Civil Rights & Discussions about civil liberties, equality, and rights of individuals or groups. \\
\addlinespace
Defense \& International Affairs & Issues about national defense, security, and international relations. \\
\addlinespace
Economy, Trade \& Labor & Issues related to economic policy, employment, labor markets, and trade agreements. \\
\addlinespace
Government Operations \& Administration & Posts related to the functioning, organization, or administration of government. \\
\addlinespace
Infrastructure \& Environment & Matters relating to public infrastructure, transportation, and environmental protection. \\
\addlinespace
Law, Crime \& Justice & Topics focused on legal systems, crime, law enforcement, and judicial matters. \\
\addlinespace
Science, Technology \& Energy & Content involving scientific research, technological development, and energy policy. \\
\addlinespace
Social Policy & Topics concerning welfare, health, education, and other social services. \\
\addlinespace
Non-Political & Content not related to politics or policy issues. \\
\bottomrule
\end{tabular}
\label{themetable}
\end{table}

\clearpage

\section{Semantic clusters}
% --- TABLE:    Subtopics and their classification ---
\begin{longtable}{p{4cm}p{8cm}c}
\textbf{Topic of Cluster} & \textbf{Description} & \textbf{Political} \\
\hline
\endfirsthead
\textbf{Topic of Cluster} & \textbf{Description} & \textbf{Political} \\
\hline
\endhead

Animation \& Gaming & Conversations about animation, anime, and gaming communities. & \\
Harry \& Meghan's Life & Focus on the lives of the Duke and Duchess of Sussex. & \\
Global Conflicts and Social Commentary & Perspectives on global conflicts and commentary. & \checkmark \\
Social and Political Discourse & Opinions on social and political matters. & \checkmark \\
AEW \& ROH Wrestling & Discussions on AEW and ROH wrestling. & \\
TikTok Ban Controversy & Debates surrounding the TikTok ban. & \checkmark \\
Georgia Protests & Coverage of protests against the Georgian government. & \checkmark \\
Jimmy Carter's Legacy & Reflections on Jimmy Carter’s life and presidency. & \checkmark \\
Loss and Reflection & Themes of longing, memory, and personal reflection. & \\
Black History \& Women's Achievements & Celebrating Black history and women's contributions. & \checkmark \\
Sports News \& Analysis & Updates and opinions on sports events. & \\
Department of Education Debate & Debates about U.S. Department of Education policies. & \checkmark \\
Aviation Safety Concerns & Plane crashes and safety issues in aviation. & \checkmark \\
Global Concerns \& Political Shifts & Interconnected political and global issues. & \checkmark \\
Political Discontent \& Critique & Dissatisfaction with politics and leadership. & \checkmark \\
Pope Francis' Death \& Vance Criticism & Reactions to Pope Francis' death and Vance criticism. & \checkmark \\
Urban Mobility \& Alternatives & Calls for improved city transport and alternatives. & \checkmark \\
Bsky Social Mentions & Mentions of the Bsky social platform. & \checkmark \\
Valentine's Day Reflections & Thoughts and feelings about Valentine's Day. & \\
Distorted Christianity & Critique of misuse of Christian beliefs. & \checkmark \\
ABDL Lifestyle & Discussion of the ABDL (Adult Baby/Diaper Lover) community. & \\
Pets \& Their Quirks & Everyday stories about pets and their behaviors. & \\
AI Concerns and Backlash & Worries and criticisms about AI technology. & \checkmark \\
Online Promotions \& Links & Promotional and advertising content online. & \\
DEI Controversy and Shifts & Debates over diversity, equity, and inclusion. & \checkmark \\
Birthday Celebrations & Posts about birthday wishes and celebrations. & \\
Holiday Celebrations & Holiday greetings and celebrations. & \\
Hegseth Security Breach & Reports on a major security incident involving Hegseth. & \checkmark \\
Economic Boycott & Advocacy for boycotting major corporations. & \checkmark \\
Southern California Wildfires & News on wildfires affecting Los Angeles and San Diego. & \checkmark \\
Federal Worker Purges & Reports on Trump administration firing federal workers. & \checkmark \\
Trump Criticism \& MAGA & Critiques of Trump and MAGA supporters. & \checkmark \\
Sonic Universe \& Fan Creations & Showcasing Sonic fan art and creations. & \\
Expressions of Gratitude & Expressions of thanks and appreciation. & \\
Executive Power \& Immigration & Concerns over executive action and immigration. & \checkmark \\
Labour and Reform Challenges & Critiques of Labour Party policies and reforms. & \checkmark \\
Research \& Collaboration & Insights into research and academic collaboration. & \checkmark \\
Sexual Content \& Fetishes & Explicit sexual topics and fetishes. & \\
Political Commentary \& Analysis & Substack posts with political commentary. & \checkmark \\
Wealth Inequality \& Critique & Frustration toward economic inequality and the wealthy. & \checkmark \\
Racial Dynamics \& Power & Discussions on race and power dynamics. & \checkmark \\
US-Greenland Tensions & Escalating tensions between the U.S. and Greenland. & \checkmark \\
Downvote/Agreement Signals & Use of online agreement or downvote signals. & \checkmark \\
University Funding \& Academic Freedom & Pressures on university funding and independence. & \checkmark \\
Public Health Concerns & Anxiety around health risks and infectious diseases. & \checkmark \\
Diverse Online Expressions & Wide variety of online opinions and content. & \\
Ukraine-Russia Conflict & Ongoing discussion of the Ukraine-Russia war. & \checkmark \\
Resistance \& Protests & Documentation of public protests and resistance. & \checkmark \\
Democratic Party Frustration & Discontent with the Democratic Party. & \checkmark \\
Navigating Life's Challenges & Stories of resilience and personal struggle. & \\
Trans Rights Under Attack & Focus on attacks against trans rights. & \checkmark \\
Birdwatching \& Photography & Sharing birdwatching experiences and photos. & \\
Cultural \& Political Commentary & Takes on cultural and political topics. & \checkmark \\
Artist Showcase \& Introductions & Artists sharing their work and introductions. & \\
Romantic Courtship & Posts about relationships and courtship. & \\
US-Canada Trade Tensions & Trade disputes between the U.S. and Canada. & \checkmark \\
Rising Egg Prices & Complaints about egg price increases. & \checkmark \\
Gaming \& Streaming Community & Twitch and streaming community discussions. & \\
Online Fiction \& Comics & Sharing online fiction and comics. & \\
Cosmic Exploration \& Observation & Astronomy and cosmic exploration topics. & \checkmark \\
Nature's Beauty & Photos and appreciation of nature scenes. & \\
Game Development Trends & Developments in the game industry. & \\
Outrage and Disgust & Expressions of outrage and disgust. & \\
Internet Humor \& Laughter & Sharing online humor and jokes. & \\
Musk's Controversies \& Government Impact & Debates on Elon Musk and government actions. & \checkmark \\
Medicaid Cuts and Social Safety Nets & Critique of Medicaid cuts and social support. & \checkmark \\
Positive Appreciation \& Praise & Expressions of positivity and praise. & \\
Confirmation and Agreement & Posts expressing strong agreement. & \checkmark \\
Furry Fursuit Community & Furry fandom and fursuit community content. & \\
Commissioned Art & Artists sharing commissioned works. & \\
Furry Art \& Vore & Furry and vore-related art. & \\
Music Discovery \& Appreciation & Discovering and sharing music. & \\
Online Video Sharing & YouTube and video sharing platform discussions. & \checkmark \\
Spooky \& Queer Fantasy & Spooky fiction and queer fantasy content. & \\
Artistic Sketches and Creations & Showcasing sketches and creative artwork. & \\
Online Shops \& Restocks & Updates on online store restocks. & \\
Commission Openings & Artists announcing commission availability. & \\
Bsky Community Networking & Users networking and sharing on Bluesky. & \\
Bluesky Migration & Bluesky as a new social media alternative. & \checkmark \\
Financial Aid \& Mutual Support & Requests and offers of mutual financial help. & \checkmark \\
Gaza Humanitarian Crisis & Coverage of the Gaza humanitarian emergency. & \checkmark \\
\label{appendix-subtopics}
\end{longtable}

\clearpage

\section{Data Collection}
\label{appendix-data-collection}

Our data collection directly received a continuous stream of unfiltered events from the Firehose API, interrupted only by two shutdowns. The first shutdown happened on 16-01-2025 and caused data loss from 16 UTC until the end of the day.

The second shutdown happened on 31-03-2025, and has caused data loss from 11 UTC to the end of the same day, as well as data loss for the entire UTC days of 01-04-2025 and 02-04-2025.

In total, we have lost 69 hours of events over the entire data collection, which corresponds to an uptime above $98\%$.

\section{Activity plots}
\begin{figure}[h]
    \centering
    \includegraphics[width=\textwidth]{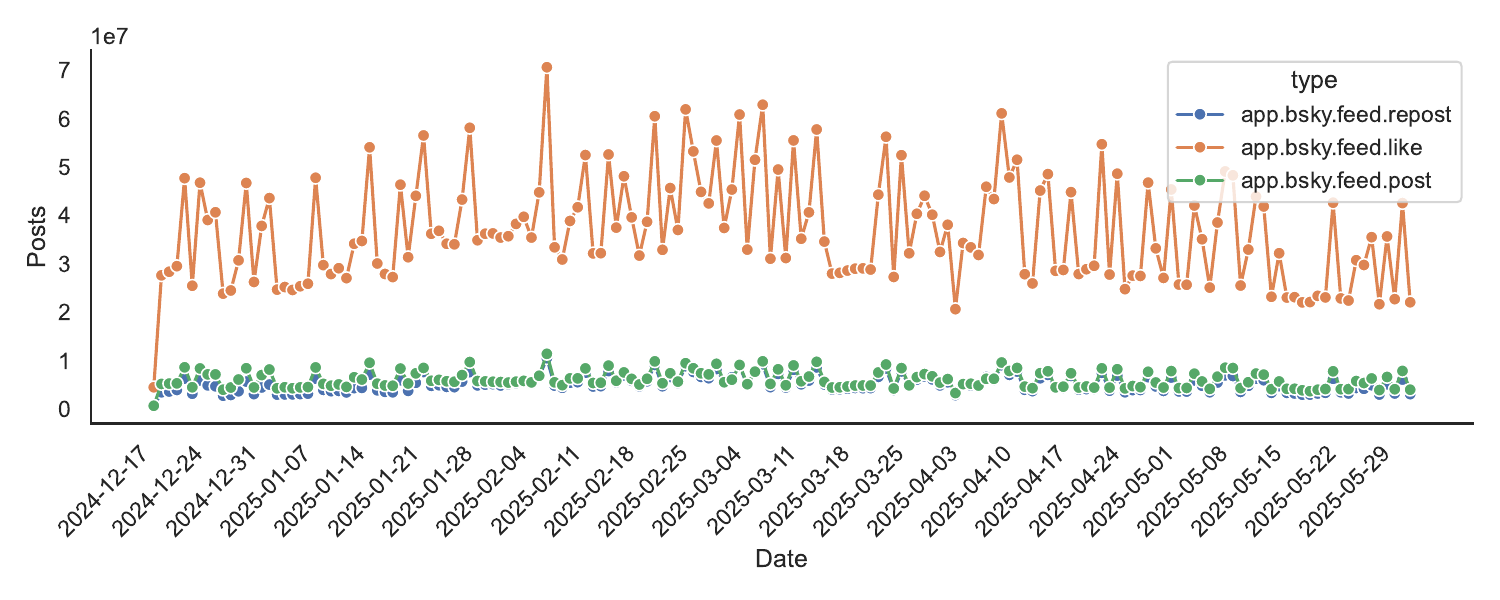}
    \caption{Number of reposts, posts, and likes in time.}
    \label{fig:posts_activities_per_day}
\end{figure}

\begin{figure}[h]
    \centering
    \includegraphics[width=\textwidth]{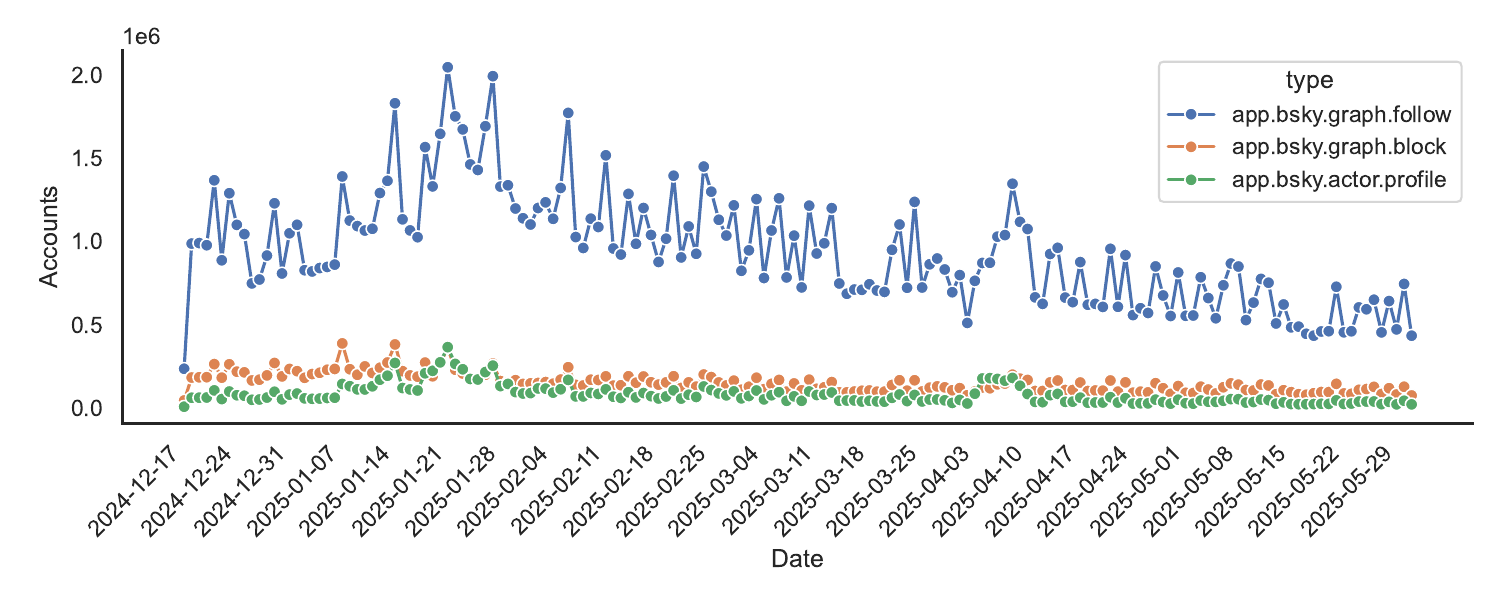}
    \caption{Number of blocks, follows, and profile creations in time.}
    \label{fig:users_activities_per_day}
\end{figure}

\clearpage

\section{Polarization between structure-based groups}
\label{appendix-aei-matrix}

\begin{figure}[htbp]
    \centering
    \begin{subfigure}[t]{0.49\textwidth}
        \centering
        \includegraphics[width=\textwidth]{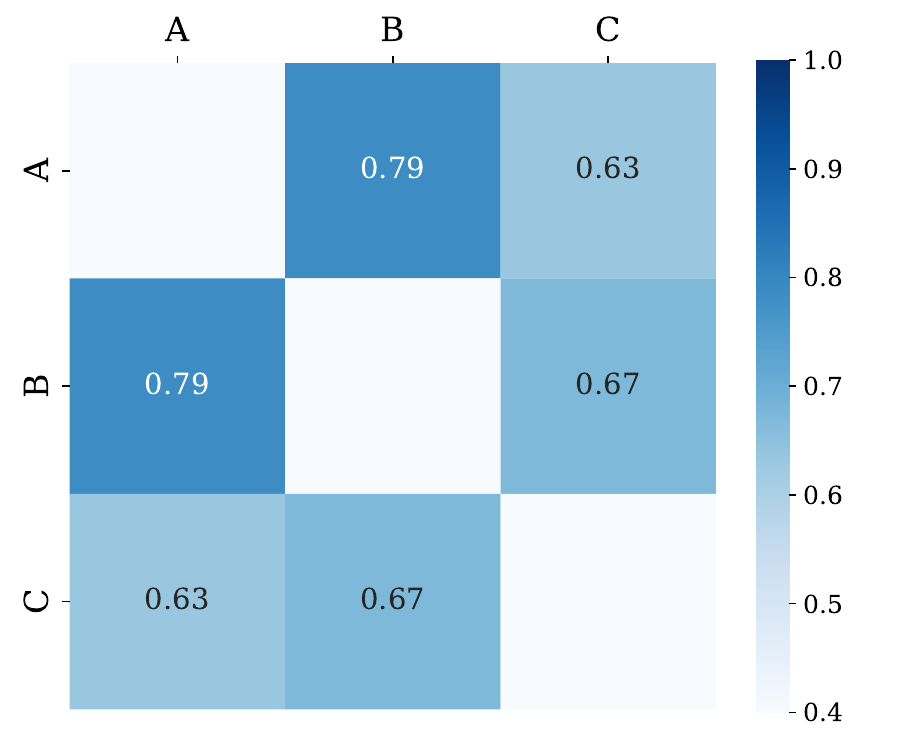}
        \caption{Trump Adminstration network}
        \label{fig:subplot1}
    \end{subfigure}
    \hfill
    \begin{subfigure}[t]{0.49\textwidth}
        \centering
        \includegraphics[width=\textwidth]{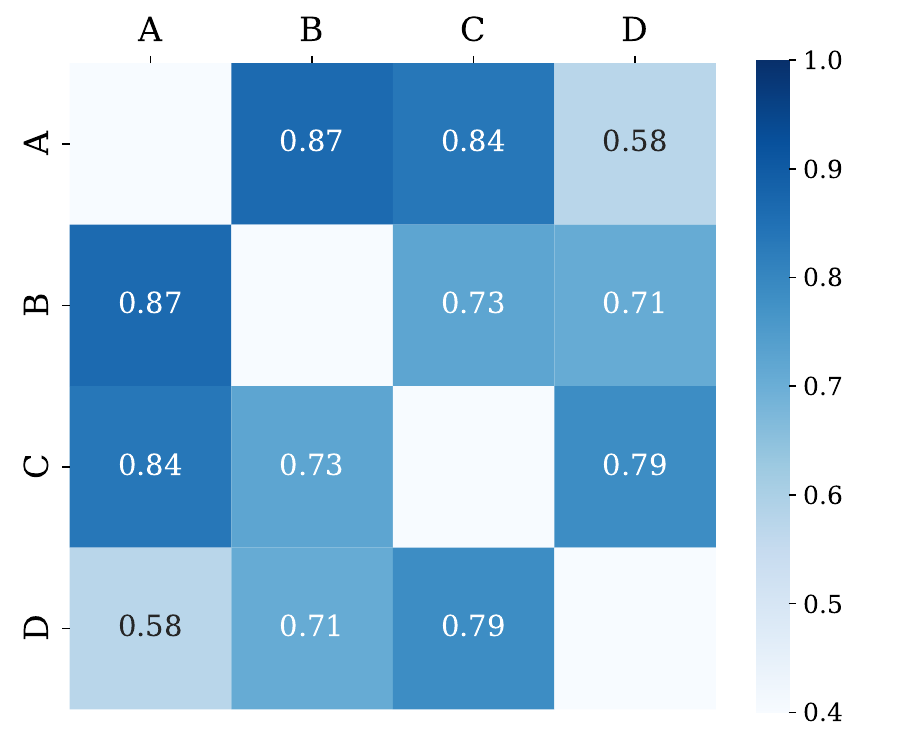}
        \caption{Elon Musk network}
        \label{fig:subplot2}
    \end{subfigure}
    \caption{Heatmap of AEI polarization scores between structural groups (labeled A--C and A--D) for the networks representing communication patterns on contemporary US politics. Higher values indicate greater division between group pairs.}
    \label{fig:mainfigure}
\end{figure}

\clearpage

\section{Network visualizations}
\label{appendix-rusukr-graph}

\begin{figure}[h]
    \centering
    \begin{overpic}[width=\textwidth]{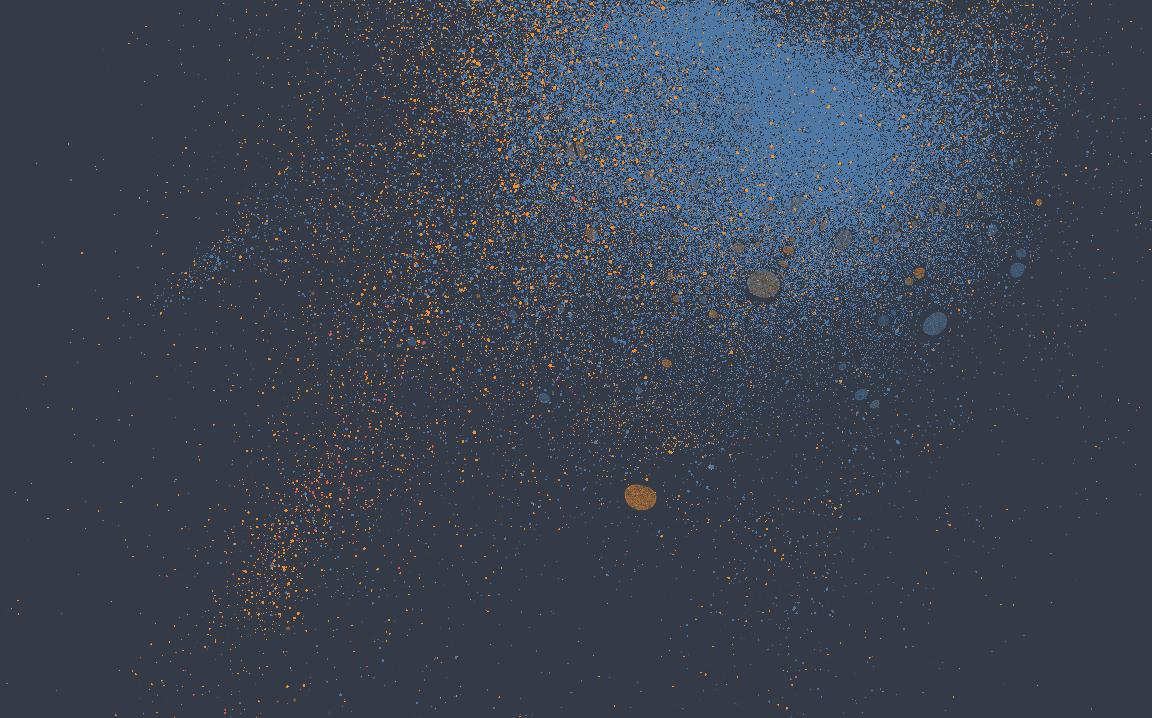}
        % Place the overlay image at (x,y) coordinates (in % of width and height)
        \put(60,0){\setlength{\fboxrule}{0.1pt}\setlength{\fboxsep}{0pt}\fcolorbox{white}{white}{\includegraphics[width=0.4\textwidth]{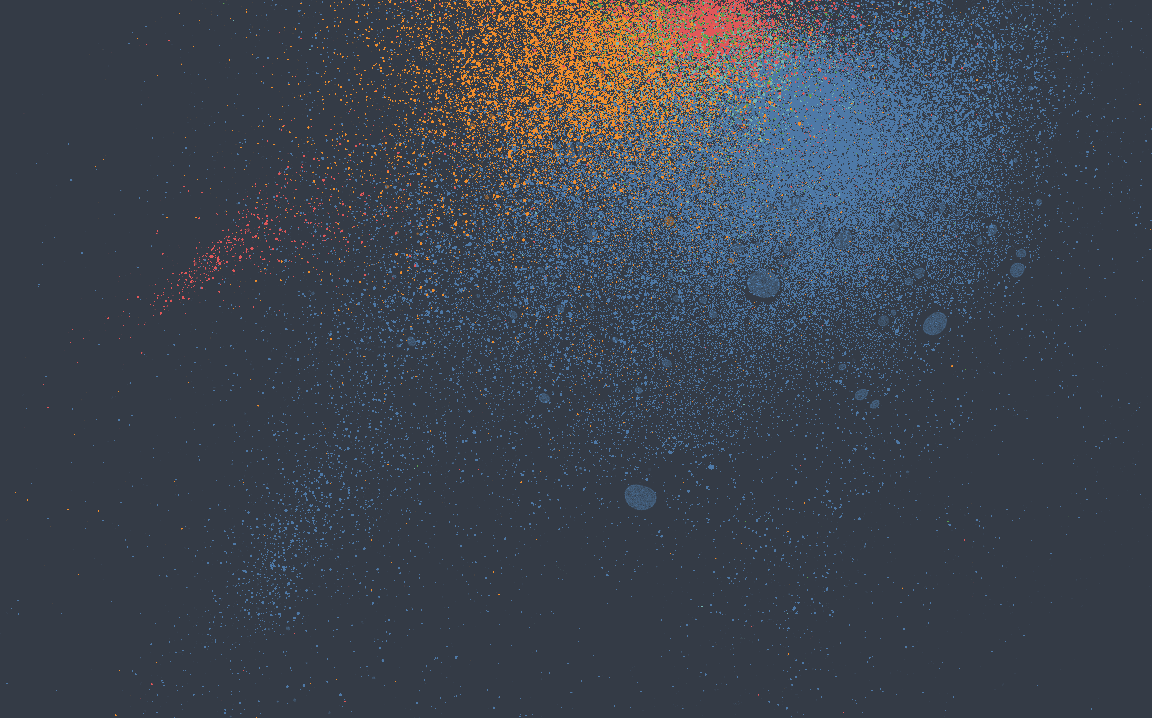}}}
    \end{overpic}
    \caption{Repost network of Russia-Ukraine. In the main plot, nodes are colored according to stance annotations inferred using the content-based approach described in Section \ref{content-based-detection} (blue: pro-Ukraine, orange: neutral, red: pro-Russia). The inset in the lower right shows the same network, with nodes colored by structural groups detected via a stochastic block model (see Section \ref{structural-based-detection}). For clarity, nodes with only one connection are omitted from both visualizations.}
\end{figure}

\clearpage

\section{Joint probability tables}
\label{appendix-joint-probs-tables}

\begin{figure}[htbp]
    \centering
    \begin{subfigure}[t]{0.49\textwidth}
        \centering
        \includegraphics[width=\textwidth]{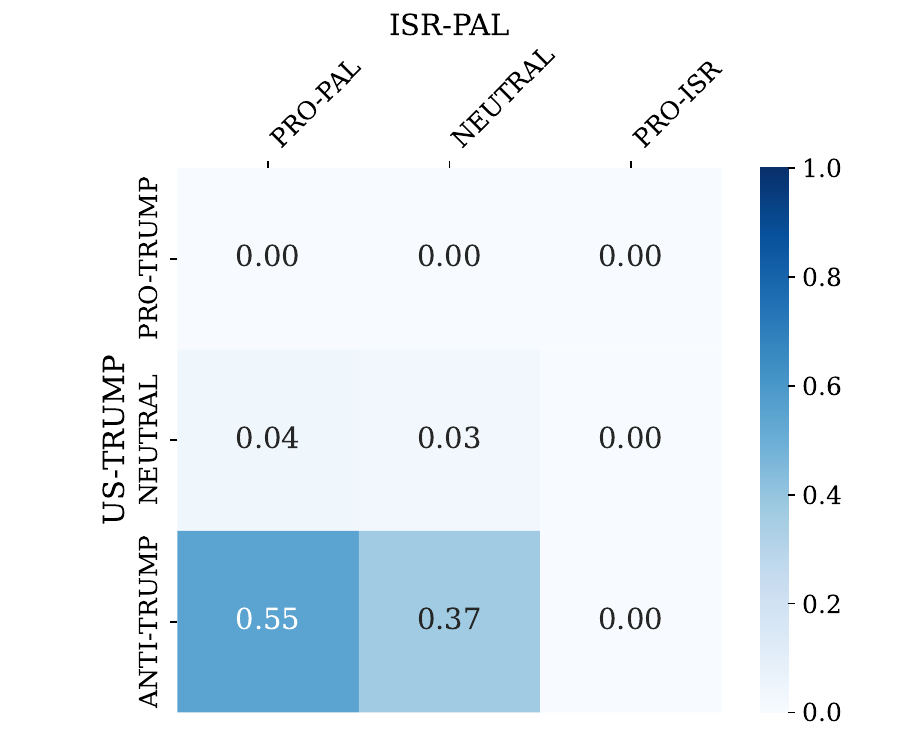}
        \caption{Israel-Palestine \& US-Trump}
        \label{fig:subplot1}
    \end{subfigure}
    \hfill
    \begin{subfigure}[t]{0.49\textwidth}
        \centering
        \includegraphics[width=\textwidth]{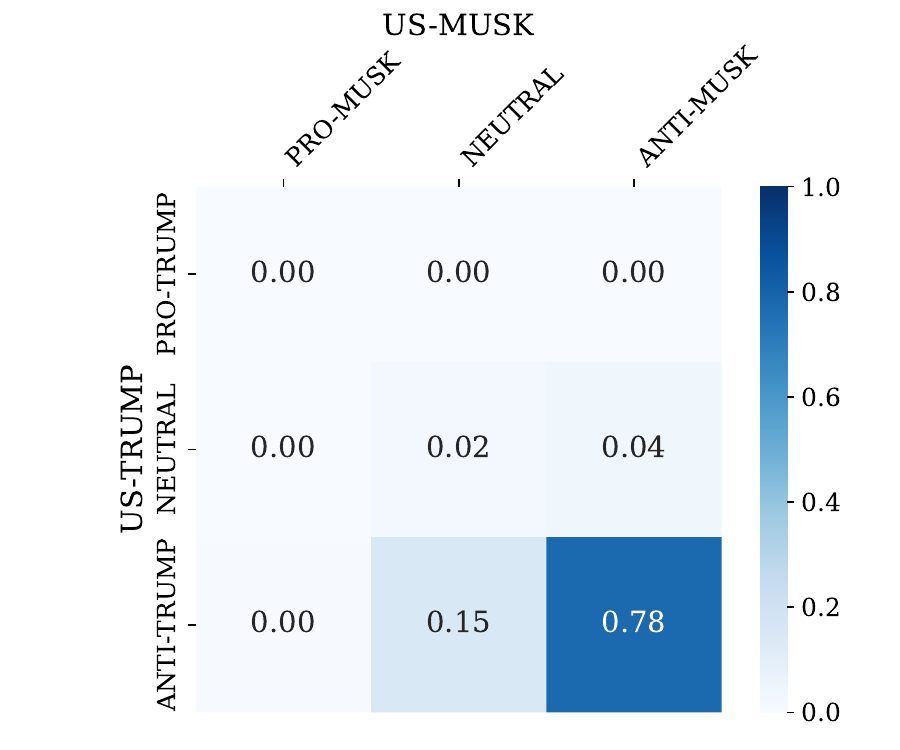}
        \caption{US-Musk \& US-Trump}
        \label{fig:subplot2}
    \end{subfigure}
    \caption{Joint probability table for joint political stances. Each cell shows the proportion of individuals taking the corresponding combination of stances on both given topics.}
    \label{fig:appendix-joint-probs1}
\end{figure}

\begin{figure}[htbp]
    \centering
    \begin{subfigure}[t]{0.49\textwidth}
        \centering
        \includegraphics[width=\textwidth]{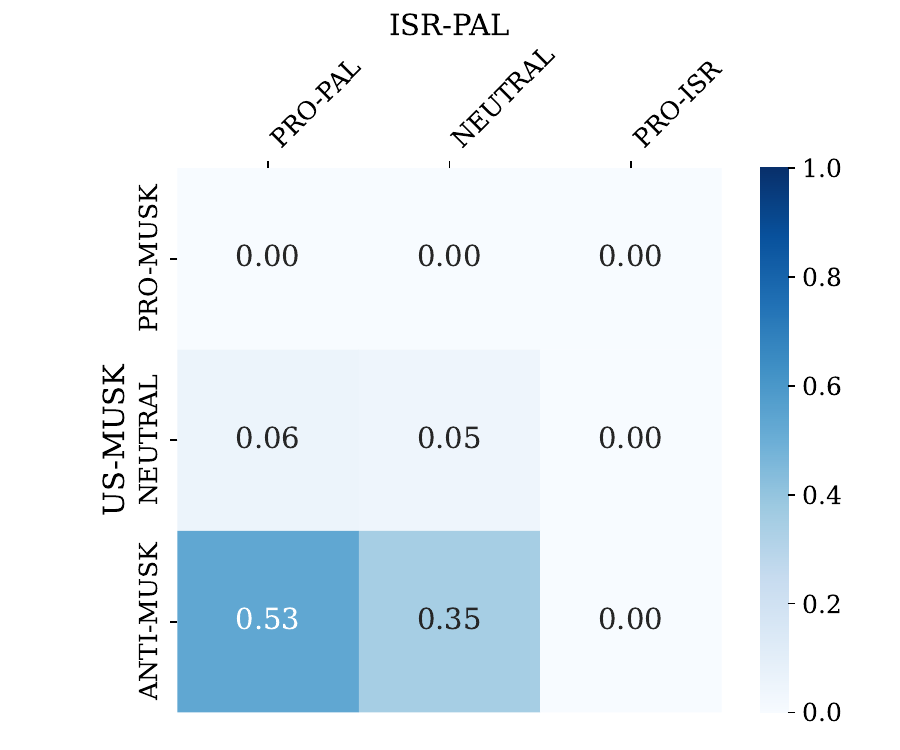}
        \caption{Israel-Palestine \& US-Musk}
        \label{fig:subplot1}
    \end{subfigure}
    \hfill
    \begin{subfigure}[t]{0.49\textwidth}
        \centering
        \includegraphics[width=\textwidth]{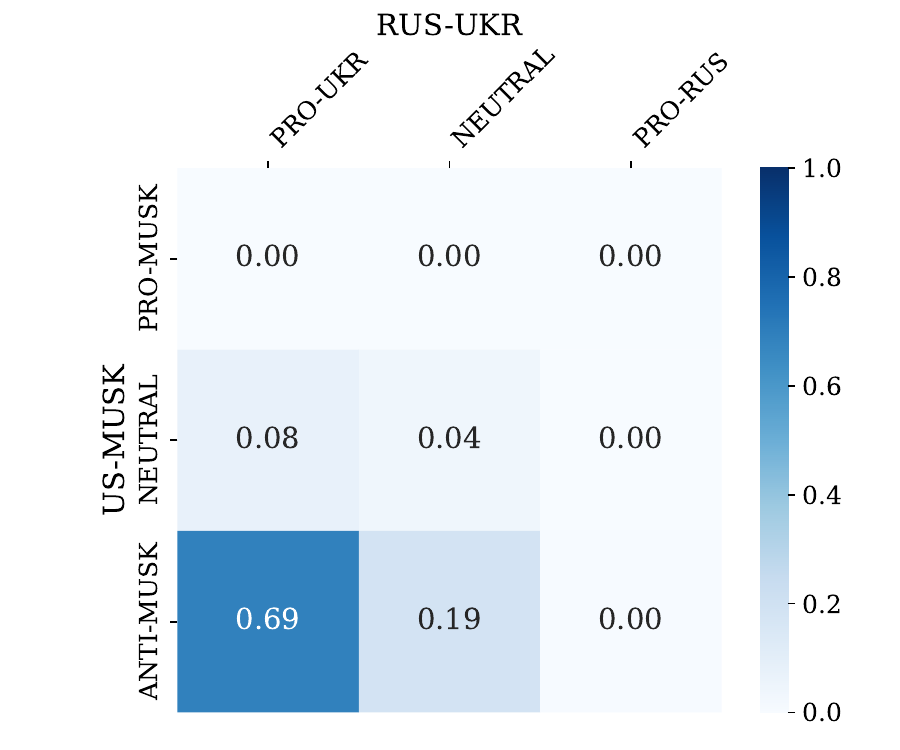}
        \caption{Russia-Ukraine \& US-Musk}
        \label{fig:subplot2}
    \end{subfigure}
    \caption{Joint probability table for joint political stances. Each cell shows the proportion of individuals taking the corresponding combination of stances on both given topics.}
    \label{fig:appendix-joint-probs2}
\end{figure}

\clearpage

%\section{Data collection}

%\begin{table}[h]
%    \centering
%    \caption{Overall Bluesky activity recorded during our data collection for \textit{create} actions of the selected types included in the analysis.}
%    \begin{tabular}{ c c c }
%        \hline
%        \textbf{Action Type} & \textbf{Total Actions} & \textbf{Total Authors} \\
%        \hline
%        Likes & 6\,032\,519\,995 &	306\,199\,225\\
%        Posts & 1\,021\,268\,020 &	164\,939\,388\\
%        Reposts & 866\,676\,384 & 106\,568\,078\\
%        Blocks & -- & 25\,408\,088\\
%        Follows & -- & 155\,281\,833\\
%        Sign-ups & -- & 13\,500\,391 \\
%        \hline
%    \end{tabular}
%    
%    \label{tab:global_stats}
%\end{table}

\end{document}